\documentclass[oneside,11pt]{article}

\usepackage[english]{babel}    
\usepackage{tgpagella}
\usepackage{enumerate} 
\usepackage{graphics,latexsym,amsfonts}       
\usepackage{amssymb,amsthm,hyperref,mathtools}   
\usepackage[dvipsnames]{xcolor} 
\usepackage{mathrsfs} 
\usepackage{graphicx} 
\usepackage{float} 
\usepackage{ulem}    
\usepackage{picture}
\hypersetup{
    colorlinks, 
    linkcolor={red!50!black},  
    citecolor={blue!50!black},  
    urlcolor={blue!80!black}   
}
\usepackage[longnamesfirst]{natbib}    
\usepackage{tabularx}

\def\reference#1{\href{#1}{Cliquer ici pour voir une r\'ef\'erence.}} 
\usepackage{sectsty} 

\usepackage{setspace}
\usepackage{caption}
\sectionfont{\normalfont\scshape\centering}
\subsectionfont{\centering}
\providecommand{\U}[1]{\protect \rule{.1in}{.1in}}
{\normalfont\itshape}

\usepackage[top=1.25in,bottom=1.25in,left=1.25in]{geometry}
\def\reference#1{\href{#1}{Click to see a reference}} 

{\color{Maroon}}

\def\X{\mathscr{X}}
\def\es{\varnothing}

\def\sbs{\subseteq}

\DeclareMathOperator*{\argmax}{argmax} 
\makeatletter

\newtheoremstyle{mytheoremstyle} 
    {\topsep}                    
    {\topsep}                    
    {\itshape}                   
    {}                           
    {\scshape}                   
    {.}                          
    {.5em}                       
    {}  

\theoremstyle{mytheoremstyle}
\newtheorem{theorem}{Theorem} 
\newtheorem*{theorem*}{Theorem}
\newtheorem{proposition}{Proposition}
\newtheorem{lemma}{Lemma}
\newtheorem{corollary}{Corollary} 
\newtheorem*{corollary*}{Corollary} 
\newtheoremstyle{mydefinitionstyle} 
    {\topsep}                    
    {\topsep}                    
    {}                   
    {}                           
    {\scshape}                   
    {.}                          
    {.5em}                       
    {}  
\theoremstyle{mydefinitionstyle}
\newtheorem{definition}{Definition} 
\newtheorem{example}{Example} 

\newtheorem*{question*}{Question}

\makeatletter

\title{Identification of consideration sets from choice data\thanks{The authors wish to thank Alfio Giarlotta, M.\,Ali Khan, and Marco Mariotti for several comments and suggestions.
Angelo Petralia acknowledges the support of "Ministero del Ministero dell'Istruzione, dell'Universit\`a e della Ricerca (MIUR), PE9 GRINS "Spoke 8", project \textit{Growing, Resilient, INclusive, and Sustainable}, CUP E63C22002120006.
Additional acknowledgments will be mentioned in the final draft.} }

\author{ 
\textsc{Davide Carpentiere}\thanks{University of Catania, Catania, Italy. davide.carpentiere@phd.unict.it}, \textsc{Angelo Petralia}\thanks{University of Catania, Catania, Italy. angelo.petralia@unict.it}   
}

\usepackage{fancyhdr} 
\date{}
\begin{document} 
\maketitle

\begin{abstract}
\noindent 
We show that many bounded rationality patterns of choice can be alternatively represented as testable models of limited consideration, and we elicit the features of the associated unobserved consideration sets from the observed choice.
Moreover, we characterize some testable choice  procedures in which the DM considers as few alternatives as possible.
These properties, compatible with the empirical evidence, allow the experimenter to uniquely infer the DM's unobserved consideration sets from irrational features of the observed behavior.
\end{abstract}

\medskip


\medskip


\section*{Introduction}\label{SECT:Introduction}

Limited consideration in individual choice has been extensively analyzed in economics, marketing, and psychology.
According to this paradigm, the DM does not examines all available alternatives, but she selects the preferred item among those that fall under her consideration.
In economic theory this phenomenon has been originally described by \cite{EliazSpiegler2011}, and formalized, among the others, by    	\cite{ManziniMariotti2014}, \cite{DemuynckSeel2018}, and \cite{GengOzbay2021}.
 A main issue of these frameworks is the identification of the collections of considered items (the \textit{consideration sets}) from data, and the elicitation of their characteristics.
This task is problematic for at least two reasons.
First, there are choice behaviors that, even if compatible with some general models of limited consideration, lack the identification of the associated consideration sets and their specific features. 
Second, whatever approach the experimenter adopts to explain an observed choice, often it is impossible to retrieve from the subject's behavior a unique collection of consideration sets.

In this work we first propose a testable notion of limited consideration that encloses many bounded rationality models discussed in the mainstream economic literature.
Under this definition it is possible to provide an alternative representation for each analyzed bounded rationality property, identify the associated family of consideration sets, and disentangle the features of  DM's consideration that can be inferred from the observed choice.
Finally, we introduce some novel models of limited consideration, assuming that DM observes the least amount of items in each menu.
Due to this constraint, it is possibile to uniquely estimates consideration sets from the analyzed choice behavior with low computational effort.

\textit{Related literature.--} First, this paper contributes to the vast literature on bounded rationality in choices.
According to the theory of revealed preferences pioneered by \cite{Samuelson1938},  individual preferences are revealed by choice data. 
In this approach rationality of choices is encoded by the notion of \textsl{rationalizability}, that is, the possibility to explain the choice behavior of a decision maker (DM) by maximizing a binary relation of revealed preference.  
However, rationalizability does not justify many observed phenomena.   
Starting from ~\cite{Simon1955}, rationalizability has been weakened by forms of \textsl{bounded rationality}, which explain  a larger portions of choices using regularity properties inspired by behavioral phenomena observed in experimental economics and psychology, such as shortlisting \citep{ManziniMariotti2012b}, limited attention \citep{MasatliogluNakajimaOzbay2012}, psychological constraints \citep{CherepanovFeddersenSandroni2013}, salience effects \citep{BordaloGennaioliShleifer2013}, and choice overload \citep{LlerasMasatliogluNakajimaOzbay2017}.
In our work we present a general but testable definition of limited consideration that allows to alternatively interpret many bounded rationality models as the observed behavior of a DM whose consideration satisfies specific features.
In doing so, we introduce several new bounded rationality approaches, which offer a complete view on the representation of limited consideration and limited attention in deterministic choice theory.   

We also provide several results about the identification of unobserved consideration sets and  their characteristics from choice datasets.
This problem has been already addressed in economic psychology and marketing, in which consideration sets are usually estimated by means of surveys \citep{PaulssenBagozzi2005,Suh2009}, and statistical inference applied on household panel data  \citep{VanNieropetal}.
In economic theory, starting from the seminal paper of \cite{ManziniMariotti2014}, many solutions have been proposed to retrieve from stochastic choice data  consideration probabilities of feasible items.
For deterministic models of choice the inference of the DM's consideration from data is still an open problem, because the same observed choice can be explained by distinct collections of consideration sets.
Our contribution to this research question is twofold: first we show that many observed choices, which are explained by models of bounded rationality, can be also justified by alternative approaches of limited consideration, characterized by consideration sets with distinguished features.
We show how to derive these features from the observed choice behavior.
Second, we introduce some models of limited consideration that describe a DM who observes few items in each menu.
This limitation on observability, compatible with the empirical evidence, allows the experimenter to identify unique consideration sets from choice data.
 
The paper is organized as follows. 
 Section~\ref{SECT:Preliminaries} contains preliminary definitions and results.
In Section~\ref{SECT:Limited_consideration} we formalize a notion of limited consideration,  which encompasses many bounded rationality properties used to explain observed choices.
 In Section~\ref{SECTION:identification_consideration_sets} we propose some  models of limited consideration that rationalize the same choice behavior justified by several bounded rationality patterns that have been investigated in the last decade, and we identify the associated
consideration sets. 
In Section~\ref{SECT:Minimal_limited_attention} we characterize some testable properties of \textit{minimal} consideration, which allows the experimenter to unambiguously deduce the DM's sets of observed alternatives from data.
All the proofs are in the Appendix.  


\section{Preliminaries}\label{SECT:Preliminaries}
In what follows, $X$ denotes the \textsl{ground set}, a finite nonempty set of alternatives.
 Any nonempty set $A \subseteq X$ is a \textsl{menu}, and $\X = 2^X \setminus \{\es\}$ is the family of all menus.
A \textsl{choice function} on $X$ is a map $c \colon \mathscr{X}\rightarrow X$ such that $c(A)\in A$ for any $A\in\X$. 
A \textsl{choice correspondence} on $X$ is a map $\Gamma\colon \mathscr{X}\rightarrow  \mathscr{X}$ such that $\Gamma(A)\subseteq A$ for any $A\in\X$.
To simplify notation, we often omit set delimiters and commas: thus, $A \cup x$ stands for $A \cup \{x\}$, $A\setminus x$ stands for $A\setminus \{x\}$, $c(xy)=x$ for $c(\{x,y\})=\{x\}$, $\Gamma(xy)=x$ for $\Gamma(\{x,y\})=\{x\}$, etc.

A binary relation $\succ$ on $X$ is \textsl{asymmetric} if $x \succ y$ implies $\neg(y \succ x)$, \textsl{transitive} if $x \succ y \succ z$ implies $x \succ z$, and \textsl{complete} if $x \neq y$ implies $x \succ y$ or $y \succ x$ (here $x,y,z$ are arbitrary elements of $X$). 
A binary relation $\succ$ on $X$ is \textsl{acyclic} if $x_1 \succ x_2 \succ \ldots \succ x_n \succ x_1$ holds for no $x_1,x_2, \ldots, x_n \in X$, with $n \geqslant 3$.
A \textsl{partial order} is an asymmetric and transitive relation.
A \textsl{weak order} is a reflexive, transitive, and complete binary relation. 
A \textsl{linear order} is an asymmetric, transitive, and complete relation.
A \textsl{strict linear order} is an irreflexive, transitive and complete relation.   
 
 Given an asymmetric relation $\succ$ on $X$ and a menu $A \in \X$, the set of \textsl{maximal} elements of $A$ is $\max(A,\succ)=\{x \in X : y \succ x \text{ for no } y \in A\}$. 
A choice function $c \colon \mathscr{X} \to X$ is \textsl{rationalizable} if there is an asymmetric relation $\succ$ on $X$ (in fact, a linear order) such that, for any $A \in \mathscr{X}$, $c(A)$ is the unique element of the set $\max(A,\succ)$; 
in this case we write $c(A) = \max(A,\succ)$.
The rationalizability of choice functions is characterized by the property of \textsl{Contraction Consistency} due to \cite{Chernoff1954}, also called \textsl{Independence of Irrelevant Alternatives} by \cite{Arrow1963}, or \textsl{Axiom}$\:\alpha$ by \cite{Sen1971}. 
This property states that if an item is chosen in a menu, then it is also chosen in any submenu containing it:
\begin{description}
	\item[Chernoff Property (Axiom$\:\alpha$):\!]
  for all $A,B\in \X$ and $x \in X$, if $x \in A \subseteq B$ and $c(B)=x$, then $c(A)=x$. 
\end{description} 
For a (finite) choice function, axiom$\:\alpha$ is equivalent to the \textsl{Weak Axiom of Revealed Preference} \citep{Samuelson1938}, which says that if an alternative $x$ is chosen when $y$ is available, then $y$ cannot be chosen when $x$ is available:
\begin{description}
	\item[WARP:\!] for all $A,B \in \X$ and $x,y \in X$, if $x,y \in A \cap B$ and $c(A) = x$, then $c(B) \neq y$.
\end{description}

Violations of axiom~$\alpha$ describe features of `irrationality'.  

\begin{definition} \label{DEF:minimal_violations_of_alpha}
	For any choice function $c \colon \X \to X$, a \textsl{switch} is an ordered pair $(A,B)$ of menus such that $A \subseteq B$ and $c(A) \neq c(B) \in A$.\footnote{\cite{CherepanovFeddersenSandroni2013} refer to such a pair of menus as \textsl{anomalous}.} 
In particular, a \textit{minimal switch} is a switch $(A,B)$ such that $\vert B\setminus A\vert=1$, or, equivalently, a pair $(A\setminus x,A)$ of menus such that $x\neq c(A) \neq c(A\setminus x)$.
\end{definition} 

A minimal switch $(A\setminus x,A)$ arises in a peculiar situation: if the DM chooses $y$ from a menu $A$, and a distinct item $x$ is removed from $A$, then the item selected from the smaller menu $A \setminus x$ is different from $y$.
As argued in \cite{GiarlottaPetraliaWatson2022b}, for a finite ground set $X$ violations of axiom $\alpha$ can always be reduced to minimal switches: 

\begin{lemma}[\citealp{GiarlottaPetraliaWatson2022b}] \label{LEMMA:minimal_violations_of_alpha}
Let $c \colon \X \to X$ be a choice function. 
For any switch $(A,B)$, there are a menu $C \in \X$ and an item $x \in X$ such that $A\subseteq C\setminus x \subseteq C \subseteq B$ and $(C\setminus x,C)$ is a minimal switch.
\end{lemma}

In view of Lemma~\ref{LEMMA:minimal_violations_of_alpha}, hereafter we shall  refer to minimal switches as switches.

%
%
%
%
%

Recall that a choice correspondence $\Gamma \colon \X \to \X$ is said to be \textsl{rationalizable} if there is an asymmetric binary relation $\succ$ on $X$ -- not necessarily a linear order -- such that $\Gamma(A)= \max(A,\succ)$ for any $A \in \X$.
In particular, $\Gamma \colon \X \to \X$ is \textsl{quasi-transitively rationalizable} if there exists a partial order $P$ on $X$ such that $\Gamma(A)= \max(A,P)$ for any $A \in \X$. 
For a finite choice correspondence, quasi-transitively rationalizability is characterized by Axioms$\:\alpha$, $\;\gamma$, and $\delta$ \citep{Sen1986}, as stated below:\footnote{This characterization fails for infinite choice correspondences, because a different property, called Axiom~$\rho$, is needed in place of Axiom~$\delta$: see \citet[Theorem 2.4 and Corollary 3.6]{CantoneGiarlottaGrecoWatson2016}.}
\begin{description}
	\item[Axiom$\:\alpha\,$:]
  for all $A,B\in \X$ and $x \in X$, if $x \in A \subseteq B$ and $x\in \Gamma(B)$, then $x\in \Gamma(A)$;\footnote{This is the natural generalization to choice correspondences of Chernoff property.}
	\item[Axiom$\:\gamma\,$:]
  for all $A,B\in \X$ and $x\in A\cap B$,  if $x\in\Gamma(A)\cap \Gamma(B)$, then $x\in\Gamma(A\cup B)$; 
	\item[Axiom$\:\delta\,$:]
  for all $A,B\in \X$ and $x,y\in X$, if $A\subseteq B$ and $x,y\in \Gamma(A)$, then $\{x\}\neq\Gamma(B)\neq\{y\}$. 
\end{description}

\section{Limited consideration}\label{SECT:Limited_consideration}
In this section we present a general definition of limited consideration, and show that this notion encompasses some bounded rationality properties of choice.
 We first introduce some preliminary concepts.

\begin{definition} \label{DEF:isomorphic_choices}
	Two choice functions $c \colon \X \to X$ and $c' \colon \X' \to X'$ are \textsl{isomorphic} if there is a bijection $\sigma \colon X \to X'$ (called an \textsl{isomorphism})  such that $\sigma(c(A)) = c'(\sigma(A))$ for any $A\in\X$.  
\end{definition}

In other words, two choice functions are isomorphic if they are equal up to a relabeling of the alternatives.

\begin{definition}\label{DEF:Property_of_choice_functions}
A \textsl{property} $\mathscr{P}$ \textsl{of choice functions} is a proper nonempty subset of the collection of all choice functions for all finite ground sets, which is closed under isomorphism.\footnote{By definition, $\mathscr P$ holds for at least one choice function on $X$, and it fails for at least one choice function on $X$.}
 This definition is in \cite{GiarlottaPetraliaWatson2022a}.
A similar notion is discussed by \cite{Lim2021} and  \cite{DeClippelRozen2023}.    
%
\end{definition}

In our framework, properties of choice functions are models of bounded rationality.
 Thus, we identify each model with the family of choice functions explained by it.
In the next two definitions we formalize the notion of consideration sets for choice functions, and we use it to propose a general yet testable paradigm of limited consideration. 
\begin{definition}
	Given a choice function $c\colon\X\to X$, a choice correspondence $\Gamma\colon\X\to\X$ is a \textsl{consideration filter for} $c$ if there is an asymmetric binary relation $\succ$ on $X$ such that $\Gamma(A)\subseteq A$, and $c(A)=\max(\Gamma(A),\succ)$ for any $A\in\X$.
	We call the pair $(\Gamma,\succ)$ an \textsl{explanation by consideration} of $c$.
\end{definition}

\begin{definition}\label{DEF:limited_consideration}
	A property $\mathscr{P}$ of choice functions is a \textsl{model of limited consideration} if there is a family $\mathscr{F}$ of choice correspondences such that
	\begin{enumerate}[\rm(i)]
	\item for each $c\colon \X\to X$ in $\mathscr{P}$, there is a consideration filter $\Gamma\colon \X\to\X$ for $c$ in $\mathscr{F}$, and 
	\item for each $c^{\,\prime} \colon \X\to X$ not in $\mathscr{P}$ there is no consideration filter $\Gamma\colon \X\to\X$ for $c^{\,\prime}$ in $\mathscr{F}$.
	\end{enumerate}
	
	
%
\end{definition}

A model of limited consideration collects all the possible choice functions that can be explained by a family of consideration filters, and can be interpreted as the observed behavior of a DM whose consideration satisfies specific features.
As showed at the end of this section, Definition~\ref{DEF:limited_consideration} is general, and encompasses many models of bounded rationality proposed in the literature.
However, note that condition (ii) of this definition implies that family $\mathscr{F}$ associated to each property $\mathscr{P}$ must have some restrictions.
Indeed, to show that $\mathscr{P}$ is a model of limited consideration we need to \textit{identify} $\mathscr{F}$, disentangling the common characteristics of its consideration filters, and proving that these filters explain all and only the choice functions belonging to $\mathscr{P}$.

In what follows we recall several choice models of bounded rationality. 
To keep focus, in some cases we omit their formal description, although we mention the behavioral properties characterizing them. 
Let $c \colon \X \to X$ be a choice function. 
Then:
 	\begin{enumerate}[\rm i)]
 		\item $c$ is \textsl{rationalizable} \citep{Samuelson1938} iff axiom$\;\alpha$ holds;
	\item $c$ is \textsl{categorize-then-choose} \citep{ManziniMariotti2012b} iff \textsl{Weak WARP} holds;\footnote{\label{FN:Weak_WARP} Weak WARP: for any $A,B \in \X$ and $x,y \in X$ with $x,y \in A \subseteq B$, if $c(B)=c(xy)=x$, then $c(A)\neq y$.}
	\item $c$ is a \textsl{choice with limited attention} \citep{MasatliogluNakajimaOzbay2012} if $c(A)=\max(\Gamma(A),\rhd)$ for all $A \in \X$, where $\rhd$ is a linear order  on $X$, and $\Gamma \colon \X \to \X$ is a choice correspondence (\textit{consideration filter}) such that for any $B \in \X$ and $x \in X$, $x \notin \Gamma(B)$ implies $\Gamma(B) = \Gamma(B \setminus x)$;

	\item $c$ is \textsl{consistent with basic rationalization theory} \citep{CherepanovFeddersenSandroni2013} if there are a choice correspondence $\psi:\X\to\X$ (\textsl{psychological constraint}) satisfying Axiom $\alpha$, and an asymmetric binary relation $\succ$ on $X$ such that $c(A)=(\max(\psi(A)),\succ)$ for any $A\in\X$ if and only if Weak WARP holds;
\item $c$ is \textsl{consistent with ordered rationalization theory} \citep{CherepanovFeddersenSandroni2013} if there are a choice correspondence $\psi:\X\to\X$ (\textsl{psychological constraint}) satisfying Axiom $\alpha$, and an linear order $\rhd$ on $X$ such that $c(A)=(\max(\psi(A)),\rhd)$ for any $A\in\X$;
	\item $c$ is \textsl{list-rational} \citep{Yildiz2016} if and only if the relation of \textsl{revealed-to-follow} is asymmetric and acyclic;\footnote{{ }Formally, $x$ is revealed-to-follow $y$ if for some $A\in\X$, either (1) $x=c(A\cup y)$ and $\big[ y=c(xy)$ or $x\neq c(A)\big]$, or (2) $x\neq c(A\cup y)$ and $\big[x=c(xy)$ or $x=c(A)\big]$.}
	\item $c$ is a \textit{shortlisting} \citep{Yildiz2016}  if and only if the relation of \textsl{related to} is asymmetric and acyclic\footnote{{ }Formally, $x$ is \textsl{related to} $y$ if for some $A\in\X$ one of the following holds: (1) $x=c(A\cup y)$ and $x\neq c(A)$, (2) $y=c(A\cup x)$ and $x=(xy)$, or (3) $y\neq c(A\cup x)$, $y=c(A)$, and $y=c(xy)$.};

	\item $c$ is \textsl{overwhelming} \citep{LlerasMasatliogluNakajimaOzbay2017} if and only if \textsl{WARP under choice overload holds} if and only if Weak WARP holds;\footnote{{ }WARP under choice overload: for any $A\in\X$, there is $x\in A$ such that for any $B$ containing $x$, if $c(B)\in A$ and $c(B^{\prime})=x$ for some $B^{\prime}\supsetneq B$, then $c(B)=x$.} 
 	\item $c$ is \textsl{rationalizable by a shortlisting procedure with capacity $k$} \citep{GengOzbay2021} if $c(A)=\max\left(\Gamma_{P}^{k}(A),\rhd\right)$ for any $A\in \X$, where $k\in\mathbb{N}$, $P$ is an asymmetric binary relation on $X$,  $\rhd$ is a linear order on $X$, and $\Gamma_{P}^{k}\colon \X\to \X$ is a choice correspondence (\textsl{shortlisting procedure with capacity $k$}) such that either $\Gamma_{P}^{k}(A)=A$ if $\vert A\vert\leq k$, or $\Gamma(A)_{P}^{k}=\max(A,P)$ if $\vert A\vert > k$, and $0\leq \vert \max(A,P)\vert\leq k$ if $\vert A\vert >k;$
 	\item $c$ admits a  \textsl{conspicuity based endogenous reference representation} \citep{KibrisMasatliogluSuleymanov2021} iff \textsl{Single Reversal Axiom}  holds;\footnote{{}Single Reversal Axiom: for every $S,T\in \X$ and distinct $x,y\in X$ such that $x,y,\in S\cap T$, if $x\neq c(S)\neq c(S\setminus x)$, then either $c(T)=y$ or $c(T\setminus y)=c(T)$.}
 	\item $c$ has a \textsl{general temptation representation}	\citep{RavidStevenson2021}  if and only if \textsl{Axiom of Revealed Temptation } holds;\footnote{{}Axiom of Revealed Temptation: for any $A\in\X$, there is $x\in A$ such that WARP holds on the collection $\{B \subseteq A : x\in B\}$} 
 	 \item $c$ is a \textsl{limited consideration model with capacity $k$} \citep{Geng2022} if and only if Weak WARP holds;
 	\end{enumerate}

Any of the listed testable properties can be represented as a model of limited consideration.
Indeed, we have:
  	
\begin{theorem}
\label{THM:bounded_rationality_models_property_limited_consideration}
	Properties i)-xii) are models of limited consideration.
\end{theorem}

In Section~\ref{SECTION:identification_consideration_sets} we discuss four models of limited consideration, and we prove that each one is equivalent, in terms of observed behavior, to some of the properties listed above. 
 Such results will enable us to complete the proof of Theorem~\ref{THM:bounded_rationality_models_property_limited_consideration}, and to elicit, for each bounded rationality model, the features of the DM's consideration sets from choice data.\footnote{As a consequence, in the Appendix the proof of Theorem \ref{THM:bounded_rationality_models_property_limited_consideration} comes after the proofs of all the results mentioned in Section~\ref{SECTION:identification_consideration_sets}.}  
 
 \section{Identification of consideration sets}
 \label{SECTION:identification_consideration_sets}

 In this section we formalize some novel declinations of limited consideration. 
In Subsection~\ref{SECTION:identification_consideration_sets}\ref{SUBSECTION:rational_limited_consideration} we propose a model of limited consideration in which observational costs matter, and we show that it is equivalent to a shortlisting procedure proposed by \cite{Yildiz2016}. 
In Subsection~\ref{SECTION:identification_consideration_sets}\ref{SUBSECT:Salient_limited_attention} the DM's attention is affected only by the most salient items she faces.
Such framework justifies the same choice behavior explained by \cite{KibrisMasatliogluSuleymanov2021} and \cite{RavidStevenson2021}.
Moreover, we will show that  rationalizability \citep{Samuelson1938} can be alternatively represented as a model of limited consideration in which the DM's attention is affected only by the most preferred item. 
Finally, in the setting analyzed in Subsection~\ref{SECTION:identification_consideration_sets}\ref{SUBSECT:Competitive_limited_attention} items pairwise compete for the DM's attention.
We prove that such approach is equivalent to the notion of \textit{list-rational choice} prosed by \cite{Yildiz2016}.         
For each of these models of limited consideration we display the characteristics of the DM's consideration sets that can be gathered from the observed violations of the weak axiom of revealed preference.

\subsection{Rational limited consideration}\label{SUBSECTION:rational_limited_consideration}

In some situations the DM is not able to evaluate all the available alternatives, but she can select some items to consider, on the base of the cost of observing such alternatives.
This phenomenon has been originally analyzed by \cite{RobertsLattin1991}, who develop a model of consideration set composition in which the DM compensates the expected benefits and costs of consideration.     
Rational consideration is a facet a more general behavioral bias, the so-called  \textsl{rational inattention} \citep{Sims2003}, which is  the tendency of the DM's to deliberately process only part of the available information needed to take a decision.
Recently, \cite{MatejkaMckay2015}, \cite{CaplinDean2015}, and \cite{CaplinDeanLeahy2019} elaborate models of optimal costly information acquisition that allow the experimenter to retrieve information costs and consideration sets from choice data.   

In our simple framework, the DM's concerns about information acquisition are included into a general consideration (or observational) cost, which affects the first step of the decision process. 
Specifically, we assume that the DM first discards all the alternatives whose observational cost is too high, and then she selects the preferred item among those she focuses on.

\begin{definition}\label{DEF:rational_limited_consideration}
	A choice function $c\colon \X\to X$ is \textit{with rational limited consideration (RLC)}  if $c(A)=\max(\Gamma^{\,\mathfrak{c}}(A),\rhd)$ for any $A\in\X$, where
	\begin{enumerate}[\rm(i)]
		\item $\rhd$ is a linear order on $X$, and
	\item $\Gamma^{\,\mathfrak{c}}\colon \X \to \X$ is a choice correspondence satisfying Axioms $\alpha$, $\gamma$, and $\delta$ (\textsl{consideration cost}).
	\end{enumerate}
\end{definition}

We assume that DM can order some alternatives according to their observational cost, but in some situations she may fail to pairwise compare some items.
As a consequence, if an item is among the cheapest to observe in a menu, then the same will happen in any submenu.
This explains why the DM's consideration, described by $\Gamma^{\,\mathfrak{c}}$, satisfies Axiom $\alpha$.
Moreover, if a subset of items is considered in two distinct menus, because its elements are the cheapest to observe in both situations, then that set must be considered in the unions of those menus.
Thus, $\Gamma^{\,\mathfrak{c}}$ satisfies Axiom $\gamma$.
Finally, if two items $x$ and $y$ are considered in $A$, for any menu $B$ containing $A$ one of the following is true: (i) $x,y$ are both observed in $B$, (ii) $x,y$ are not observed in $B$, (iii) $x$ and some other item $z\neq y$ is observed in $B$, or (iv) $y$ and some other item $z\neq x$ is observed in $B$.
If observing $x$ and $y$ in $A$ is convenient, then the same must happen in any $B$ including $A$, unless a new item $z$, whose observational cost is lower than that of $x$ or $y$, is added to $A$.
If this assumption is true, then Axiom $\delta$ must hold for $\Gamma^{\,\mathfrak{c}}$. 
   
This behavioral pattern explains the same choice behavior justified by the model of  \cite{Yildiz2016}, in which the DM selects the alternative that survives to the sequential applications of two distinct decision criteria.

\begin{definition}[\citealp{Yildiz2016}]\label{DEF:shortlisting}
	A choice function $c\colon\X\to\X$ is a \textit{shortlisting (SL)} if there are a partial order $\succ$ and a linear order $\rhd$ on $X$ such that, for all $A \sbs X$, we have $c(A)=\max(\max(A, P), \rhd)$.
\end{definition}

Indeed, we have:\footnote{\,The proof of Lemma~\ref{LEM:rational_attention_equivalent_shortlisting} is straightforward, and it is left to the reader.}

\begin{lemma}\label{LEM:rational_attention_equivalent_shortlisting}
	A choice function is with rational limited consideration if and only if it is a shortlisting.
	\end{lemma}


Lemma \ref{LEM:rational_attention_equivalent_shortlisting} shows that shortlisting (property~vi) listed in Section~\ref{SECT:Limited_consideration}) is a model of limited consideration.
Moreover,  rational limited consideration allows to retrieve information on the DM's consideration cost from  the observed violations of Axiom $\alpha$.

\begin{lemma}\label{LEMMA:necessary_conditions_of_RLC}
	Let $c\colon \X\to X$ be a RLC choice function, and $(\Gamma^{\,\mathfrak{c}},\rhd)$ an explanation by consideration of $c$.
	If $x\neq c(A)\neq c(A\setminus x)$ for some $A\in\X$ and $x\in X$, then $c(A\setminus x)\rhd c(A)$, $c(A\setminus x)\not\in\Gamma^{\,\mathfrak{c}}(A)$, and $\Gamma^{\,\mathfrak{c}}(A)\supsetneq c(A)$.
\end{lemma}

Thus, for choice functions with rational limited considerations  minimal violations of rationality happen only if the removal of an item from the menus induces the DM to consider and eventually select an item that has not been evaluated before.
Specifically, the existence of a switch $(A\setminus x,A)$ implies that the cost of observing $x$ is lower than the cost of observing $c(A\setminus x)$, whereas $c(A)$ and $c(A\setminus x)$ are not comparable.
Thus, in the menu $A$ the item $x$ rules $c(A\setminus x)$ out of $\Gamma^{\,\mathfrak{c}}$, whereas in the menu $(A\setminus x)$ the alternative $c(A\setminus x)$ is considered and chosen, if it is preferred to $c(A)$.
Finally, note that the DM must consider at least one item other than $c(A)$ when she faces the menu $A$.
 Otherwise, $c(A)$ would be more convenient to observe than any other alternative in $A$, and, since the evaluation of the cost of observing items is transitive, also cheaper than $c(A\setminus x)$.
 This situation would yield $c(A\setminus x)\not\in \Gamma^{\,\mathfrak{c}}(A\setminus x)$, which is impossible. 




\subsection{Salient limited attention}\label{SUBSECT:Salient_limited_attention}

In some situations the DM does not consider some available items because he does not  notice them.
This phenomenon, known as \textit{limited attention}, 
has been empirically verified \citep{ChiangChibNarasimhan1999,GilbrideAllenby2004,Aguiaretal2023}, and it has been first formalized in choice theory by \cite{MasatliogluNakajimaOzbay2012}, and extended by \cite{Cattaneoetal2020} to a stochastic setting.
As formally described in property iii) of Section~\ref{SECT:Limited_consideration}, the DM selects an item from a menu by maximizing a linear order on the subset of elements that attract her attention.
Limited attention is a flexible property, which allows to justify a relatively large portion of choice functions defined on small ground sets \citep{GiarlottaPetraliaWatson2022a}.
Here we discuss a special case of limited attention, in which notable items in the menu affect the DM's consideration. 

Indeed, the DM's attention can be distorted by \textsl{salience} of items, usually intended as the relatively extreme position of alternatives in the DM's evaluation.
\cite{BordaloGennaioliShleifer2012,BordaloGennaioliShleifer2013} argue that attention may be captured by specific features of items (e.g., quality and price), which distort the DM's evaluation of alternatives.
Recently, \cite{Lanzani2022} proposes an axiomatization of salience theory under uncertainty.
These approaches analyze the effect of salience on the DM's judgment, but they do not elaborate on the connection between salience and limited attention.
However, new findings in psychology \citep{ParrandFriston2017,ParrandFriston2019} report that these two phenomena may interact and jointly determine individual decisions.  
Recently, \cite{GiarlottaPetraliaWatson2022b} propose a model of limited attention in which only salient items affect the DM's consideration.

\begin{definition}[\citealp{GiarlottaPetraliaWatson2022b}] \label{DEF:CSLA} \rm 
	A choice function $c \colon \X \to X$ is a \textsl{choice with salient limited attention} \textsl{(CSLA)} if $c(A) = \max(\Gamma^{\,s}(A),\rhd)$ for all $A \in \X$, where
\begin{enumerate}[\rm(i)]
	\item $\rhd$ is a linear order on $X$, and  
	\item $\Gamma^{\,s} \colon \X \to \X$ is a choice correspondence (\textsl{salient attention filter}) such that for all $B \in \X$ and $x \in X$, $x \neq \min(B,\rhd),\max(\Gamma^{\,s}(B),\rhd)$ implies $\Gamma^{\,s}(B) \setminus x = \Gamma^{\,s}(B \setminus x)$.  
\end{enumerate}
\end{definition} 

 In this specification of \cite{MasatliogluNakajimaOzbay2012} a variation in some of the DM's consideration sets occurs only if the worst option or the best observed alternative are removed from the menu.
Instead, in the general case the set of observed items can change due to the removal of any  item belonging to it.
The representation offered in Definition~\ref{DEF:CSLA} is equivalent to that of two models of bounded rationality, that discuss the effects of temptation and reference dependence on individual choices.


	\begin{definition}[\citealp{RavidStevenson2021}]
		A choice function $c\colon \X\to X $ has a \textit{general-temptation representation (GTR)} if functions $u\colon X\to \mathbb{R}$, $v \colon X\to \mathbb{R}$, and $W\colon\mathbb{R}^{2}\to\mathbb{R}_{+}$ exist such that $W$ is strictly increasing in both coordinates and $$c(A)=\arg \max_{x\in A}W(u(x), v(x)-\max_{y\in A}v(y))$$ holds for any $A\in \X$.
\end{definition}

In any menu the agent selects the item that brings her the best trade-off between her true preference, described by $u$, and the self-control cost $v$ of resisting the most tempting feasible option.
Note that in a general temptation representation the DM's consideration does not play any role.
The same is true for the following bounded rationality property.

\begin{definition}[\citealt{KibrisMasatliogluSuleymanov2021}]
	A choice function $c\colon\X\to X$ admits a \textit{conspicuity based endogenous reference representation} (\textit{CER}) if there exist a family of injective functions $U=\{U_{\rho}\}_{\rho\in X}$ and a strict linear order $\gg $ on $X$ such that 
	
	$$c(A)=\argmax_{x\in A}U_{r(A)}(A),$$

where $r(A)=\max(A,\gg)$.	 
\end{definition}

According to this procedure, in any menu $A$ the DM's perception is captured by the most conspicuous item $r(A)$, which moves her to adopt some utility function $U_{r(A)}$ in her choice.
The choice behavior explained by general temptation and conspicuity based endogenous reference representation is also justified by salient limited attention.
Indeed, we have:

\begin{lemma}
\label{LEM:CSLA_equivalent_representation}
The following are equivalent for a choice $c$:
\begin{itemize}
	\item[\rm(i)] $c$ is a choice with salient limited attention; 
	\item[\rm(ii)] $c$ has a general temptation representation;
	\item[\rm(iii)] $c$ admits a conspicuity based endogenous reference representation.
	\end{itemize}
\end{lemma}

Lemma~\ref{LEM:CSLA_equivalent_representation} show that choice data explained by temptation and conspicuity based endogenous reference representation can be interpreted also as the behavior of a DM with limited consideration.  
The features of consideration sets can be retrieved from the observed choice as follows.

\begin{lemma}[\citealt{GiarlottaPetraliaWatson2022b}]
Let $c\colon \X\to X$ be a choice with salient limited attention.
If there are $A \in \X$ and distinct $x,y\in A$ such that $x \neq c(A)\neq c(A\setminus x)$, then there is an explanation by consideration $(\Gamma^{\,s},\rhd)$ of $c$ such that $y\rhd x$, $x\in \Gamma^{\,s}(A)$, and $x$ is equal to $\min(\Gamma^{\,s}(A),\rhd)$.
\end{lemma}

In other words, for a choice with salient limited attention, if removing $x$ from a menu $A$ containing $y$ causes a switch, then we can deduce not only that the DM prefers $y$ to $x$ and pays attention to $x$ at $A$, but also that $x$ is the least preferred item among those brought to her attention in $A$.
We now single out a special case of salient limited attention, in which the  DM's consideration is affected only by the best considered item.
\begin{definition} \label{DEF:CSSLA} \rm 
	A choice function $c \colon \X \to X$ is a \textsl{choice with selective salient limited attention} \textsl{(CSSLA)} if $c(A) = \max\left(\Gamma^{\,sel}(A),\rhd\right)$ for any $A \in \X$, where
\begin{enumerate}[\rm(i)]
	\item $\rhd$ is a linear order on $X$, and   
	\item  $\Gamma^{\,sel} \colon \X \to \X$ is a choice correspondence (\textsl{selective salient attention filter}) such that for all $B \in \X$ and $x \in X$, $x \neq\max\left(\Gamma^{\,sel}(B),\rhd\right)$ implies $\Gamma(B) \setminus x = \Gamma^{\,sel}(B \setminus x)$.  
\end{enumerate}
\end{definition}

According to Definition~\ref{DEF:CSSLA}, a variation in the DM's consideration set occurs only if the best item, among those observed, is removed from the menu.
Selective salient attention identifies rationalizable choice behavior.

\begin{lemma}\label{LEM:CSSLA_alpha}
A choice function is a choice  with selective salient limited attention if and only if it satisfies Axiom $\alpha$.
\end{lemma}

The above result shows that rationalizability is a model of limited consideration.
This is not surprising, since, according to the existing literature on limited consideration, the experimenter can interpret a rationalizable choice as the observed outcome of a DM who considers in each menu all the feasible alternatives. 
However, Lemma~\ref{LEM:CSSLA_alpha} suggests that such interpretation is not the unique one.
Indeed, for any choice with selective salient limited attention, there is an explanation under limited consideration in which some available items are not observed by the DM.

\begin{lemma}
\label{LEM:CSSLA_non_trivial_consideration_sets}
	If $c\colon \X\to X$ is a choice with selective salient limited attention, then there is an explanation by consideration $(\Gamma^{\,sel},\rhd)$ of $c$ such that $\Gamma^{\,sel}(A)\subset A$ for some $A\in \X$.

\end{lemma} 

The equivalence between selective salient limited attention and rationality stems from the analysis of choice data.
In a model of limited consideration, violations of Axiom $\alpha$ occur only if an item selected in a menu is still available in a submenu, but does not belong to the DM's consideration anymore.
For choices with selective salient limited attention, variations of the selective salient attention filter are caused only by the removal of the selected item, ruling out any violation of Axiom $\alpha$.
 


\subsection{Competitive limited attention}\label{SUBSECT:Competitive_limited_attention}

     
A massive strand of the marketing literature reports a fierce competition among products and brands to gain the DM's attention, and exclude the competitors.  
Due to individual tastes, aggressive advertising, and time pressure, brands that receive attention tend to  substitute alternative brands in the consumer's consideration, until some more notable products are introduced in the market.
\cite{FaderMcAlister1990} develop and compute parametric estimations of a model in which consumers do not consider brands that are not on promotion.
\cite{AllenbyGinter1995} find that in-store merchandising increases the probability that products with the same brand names are jointly considered, and  reduces the likelihood that  alternative brands are taken into account by households.
In the work of \cite{TeruiBanAllenby2021} there is evidence that advertising increases the probability that a given item is included in the consumer's consideration set.

We propose a model of choice that reproduces the effects of marketing strategies on individual choice, by assuming that some brands and products may rule out others from the DM's consideration.
Some notation: given a linear order $\rhd$ on $X$, and $x,y\in X$ such that $y\rhd x$, denote by $^{\rhd}]x,y[$ the set $\{z\in X \colon z\rhd x \,\wedge y\rhd z \}$.
Moreover, let
\[\mathring{(xy)}_{\rhd}=
\begin{cases}
       ^{\rhd}]x,y[\;\;\;\text{if y\,$\rhd$\,x,}\\
      ^{\rhd}]y,x[ \;\;\;\text{if x\,$\rhd$\,y.}\\ 
     \end{cases}
\]

\begin{definition} \rm \label{DEF:list_attention}
	A choice function $c\colon\X\to X$ is a \textit{choice with competitive limited attention (CCLA)} if $c(A)=\max(\Gamma^{\,co}(A),\rhd)$ for any $A\in\X$, where:
	\begin{enumerate}[\rm(i)]
		\item $\rhd$ is a linear order on $X$ (\textsl{preference}), and
		\item $\Gamma^{\,co}: \X \to \X$ is a choice correspondence (\textit{competitive attention filter}), such that the following properties hold for all nonempty menus $A$:
		\begin{enumerate}
		\item if $x \in A$ and $x \notin  \Gamma^{\,co}(A)$, then $\Gamma^{\,co}(A)=\Gamma^{\,co}(A\setminus x)$;
		\item $\Gamma^{\,co}(A\setminus\min(A,\rhd))=\Gamma^{\,co}(A)\setminus\min(A,\rhd)$ or $\Gamma^{\,co}(A)=\min(A,\rhd)$;
		\item for all $x,y \in X$, if $A \subseteq\,^\rhd ]x,y[$ and $y \notin \Gamma^{\,co}(xy)$, then $y \notin \Gamma^{\,co}(A \cup xy)$.
	\end{enumerate}
	\end{enumerate}
\end{definition}

The DM selects the preferred alternative among those that capture her attention.
As for choices with limited attention, condition (ii)(a) imposes that ignored items do not influence the DM's attention when they are removed from the menu. 
Moreover, the DM's attention is never affected by the worst item, unless it is the only one that the DM observed, as described in condition (ii)(b).
Finally, condition (ii)(c) states that if an item $y$ is not  considered in presence of another item $x$ when they are the only available alternatives, then the same will happen in any menu containing $x$, $y$, and any other item that lies between $x$ and $y$ in the DM's evaluation.
The alternatives $x$ and $y$ compete for the DM's attention, and if one  overcomes the other in a pairwise comparison, then it will always shade the other in the DM's mind, as long as no more notable items (worst than $x$ or better than $y$) are introduced in the menu.
Competitive limited attention explains the same choice data justified by \textsl{list-rational} choices, discussed by \cite{Yildiz2016}.


\begin{definition}[\citealp{Yildiz2016}] \rm \label{DEF:list}
    A choice function $c$ on a set $X$ is \textit{list rational} (\textsl{LR}) if there is a linear order $\mathbf{f}$ (a \textit{list}) on $X$ such that for each menu $A \in\X$, we have $c(A)=
    c\big(\{c(A\setminus \min(A,\mathbf{f})),\min(A,\mathbf{f})\}\big)$.
\end{definition}

According to this choice procedure the DM picks in each menu the alternative that survives from a binary comparison of alternatives on a list.
List rational behavior can be equivalently interpreted by the experimenter as the observed choice of a DM whose attention is affected by marketing strategies that aim to exclude some products from her consideration.

\begin{theorem} \label{THM:list_rationalizable}
	A choice function $c$ is list rational if and only if it is a choice with competitive limited attention.
\end{theorem}

Theorem \ref{THM:list_rationalizable} shows that list rationality is a model of limited consideration.
In the long proof of this result, we investigate the connection between the model of \cite{Yildiz2016} and choice with competitive limited attention.
However, it is worth noting that the DM's list $\mathbf{f}$, which appears in Definition~\ref{DEF:list}, is identified by the preference $\rhd$ of Definition~\ref{DEF:list_attention}, as showed by Lemma~\ref{LEMMA:LR_property} in the Appendix. 
Moreover, the experimenter can retrieve from violations of Axiom $\alpha$ associated to list rational choices information about the DM's consideration sets, and her preference.
Indeed, we have:

\begin{lemma}\label{LEMMA:necessary_conditions_of_CCLA}
Let $c\colon \X\to X$ be a choice with competitive limited attention, and $(\Gamma^{\,co},\rhd)$ be some explanation by consideration of $c$. 
If there are $A \in \X$ and distinct $x,y\in A$ such that $y=c(A)\neq c(A\setminus x)$, then $x\in \Gamma^{\,co}(A)$, $x\neq \min(\Gamma^{\,co}(A),\rhd)$, and $y\rhd x$.
Moreover, if there are $x,y\in X$, and $A\in\X$ such that $y=c(xy)\neq c(A\cup xy)=x $, then either $A\subseteq \,^{\rhd}]x,y[$\,\,, $x\in\Gamma^{\,co}(xy)$, $y\rhd x$, and $y\not\in\Gamma^{\,co}(A\cup xy)$, or $A\setminus\,\mathring{(xy)}_{\rhd}\neq \varnothing$.  \end{lemma}

For a choice with competitive limited attention, if removing $x$ from a menu $A$ in which $y$ is selected causes a minimal switch, then we can infer that the DM prefers $y$ to $x$, pays attention to $x$ at $A$, but also that $x$ is not the least preferred item among those brought to her attention in $A$.
Moreover, if an item $y$ is picked in a binary comparison with $x$, but $x$ is selected when a bundle of alternatives $A$ is added to the menu $xy$, we conclude that either $y$ is preferred to $x$, $x$ is considered in $xy$ and any alternative in $A$ is preferred to $x$ but not to $y$, or, alternatively, there is at least an item in $A$ which holds an extreme position with respect to $x$ and $y$ in the DM's judgement.

%
%
Theorem \ref{THM:list_rationalizable} is the last result needed to prove Theorem \ref{THM:bounded_rationality_models_property_limited_consideration}. 
Table~\ref{TABLE:summary_limited_consideration} summarizes the relation between the bounded rationality properties listed in Section~\ref{SECT:Limited_consideration} and the models of limited consideration described in Section~\ref{SECTION:identification_consideration_sets}.  

\begin{table}[h]
\begin{center}
\begin{tabular}{|c|c|c|}
\hline
$\mathscr{P}$ & $\mathscr{P}^{\,eq}$ & $\mathscr{F}$ \\
\hline
 SL & RLC & $\Gamma^{\,\mathfrak{c}}$\\
\hline
GTR, CER & CSLA & $\Gamma^{s}$\\
\hline
Axiom $\alpha$ & CSSLA & $\Gamma^{\,sel}$\\
\hline
LR & CCLA & $\Gamma^{\,co}$ \\
\hline
\end{tabular}
\end{center}
\caption{Each bounded rationality property $\mathscr{P}$ in the first column is equivalent, in terms of observed behavior, to one of the properties $\mathscr{P}^{\,eq}$ listed in the second column, discussed in Section~\ref{SECTION:identification_consideration_sets}
In the third column we display the families of consideration filters $\mathscr{F}$ associated to each model of limited consideration.}
\label{TABLE:summary_limited_consideration}
\end{table}

In the next section we discuss some model of limited consideration in which the DM observes the least possible number of alternatives.

\section{Identification of unique consideration sets}\label{SECT:Minimal_limited_attention}

The models of limited consideration analyzed in the previous sections enable the experimenter to verify only whether some specific items in the menu are considered by the DM.
However, these methods admit \textit{multiple} explanations, and many distinct collections of consideration sets that justify the same observed choice.
Some of these explanations may exhibit a consideration \textit{larger} than the one applied by the DM in the choice.
Indeed, advances in empirical marketing and psychology \citep{HauserWernerfelt1990,RobertsLattin1991,CromptonAnkomah1993,MeissnerStraussTalluri2013} indicate that many consumers consider a surprisingly small amount of alternatives before purchasing a product.

Moreover, the identification of \textit{unique} consideration sets from the observed choice is still an open problem.
For stochastic choice functions, full identified models of limited consideration have been investigated, among the others, by \cite{ManziniMariotti2014}, \cite{BradyRehbeck2016}, \cite{Aguiaretal2023}.
 Unique consideration probabilities can be obtained also in the approach of \cite{AbaluckAdams-Prassl}, who assume that the experimenter is endowed with a richer dataset, which includes also the prices of purchased items.
For deterministic choice data, the absence of unique consideration filters is a structural problem.
This defection is even more evident for extremely general models, such as the limited attention property of \cite{MasatliogluNakajimaOzbay2012}, and
the basic rationalization theory of \cite{CherepanovFeddersenSandroni2013}, which, at least for small ground sets, explain a large portion of choices, and offer, for the same observed choice dataset, many distinct justifications.

To identify small and unique consideration sets, we reverse the perspective, and we assume \textit{ex-ante} that the DM's consideration is \textit{minimal}, i.e. the DM observes the least amount of items compatible with the features of consideration sets analyzed in \cite{MasatliogluNakajimaOzbay2012} and \cite{CherepanovFeddersenSandroni2013}.
This hyphothesis allows the experimenter to uniquely identify the DM's consideration sets from minimal irrational features of the observed choice.

\subsection{Minimal limited attention}

We first shape a choice procedure in which a DM with limited attention observes few items.
To do so, we introduce some preliminary notation.

\begin{definition}\label{DEF:CLA_switch_sets}
	Given a choice $c\colon\X\to X$ on $X$, and menu $A\in\X$,	let $S_A=\{x \in A : (A\setminus x,A) \hbox{ is a switch}\}$.
	\end{definition}

For any menu $A$, $S_A$ contains all the items whose removal causes a minimal violation of Axiom $\alpha$ for $c$.
The collection of sets $\{S_A\}_{{A}\in\X}$, and the selected items $\{c(A)\}_{A\in\X}$, bring \textsl{in nuce} all the information needed to describe the observed choice.
For choices with limited attention, this amount of information is always contained in the attention filter.

\begin{lemma}\label{LEM:minimal_attention}
If $c \colon \X \to X$ is a choice with limited attention, and $(\Gamma,\rhd)$ is an explanation by consideration of $c$, then $S_A\cup c(A)\sbs \Gamma(A)$ for any $A\in\X$.
\end{lemma}

Given Lemma~\ref{LEM:minimal_attention}, a natural task is to describe a DM affected by limited attention, but whose consideration sets are as narrow as possible.

\begin{definition}\label{DEF:minimal_attention}
	A choice function $c\colon\X\to X$ is a choice \textsl{choice with minimal limited attention (CMLA)}  if $c(A)=\max(\Gamma^{\,min}(A),\rhd)$ for any $A\in\X$, where
	\begin{enumerate}[\rm(i)]
		\item $\rhd$ is a linear order on $X$ (\textsl{preference}),
		\item $\Gamma^{\,min}:\X\to\X$ is the choice correspondence (\textsl{minimal attention filter}) such that
		\begin{itemize}
			\item[(a)]$\Gamma^{\,min}(A \setminus x)=\Gamma^{\,min}(A)$ for any $A \in\X$ and $x \notin \Gamma^{\,min}(A)$, and
			\item[(b)]  
			$\Gamma^{\,min}(A)=S_A \cup c(A)$ for any $A\in \X$.
		\end{itemize} 
	\end{enumerate}
	\end{definition}
	
	 As for the general model of limited attention, the DM selects the preferred item among those observed.
	Condition (ii)(a) of Definition~\ref{DEF:minimal_attention} is typical of choices with limited attention, and requires that the removal of unobserved items from the menu does not modify the DM's attention filter.  
	Moreover, according to condition (ii)(b) the DM's consideration is \textit{minimal}, because in any menu she observes the minimum number of items admitted by the limited attention model of \cite{MasatliogluNakajimaOzbay2012}.  
	Choices with minimal limited attention replicate the evidence documenting that consumers observes the fewest possible number of alternatives from each menu.
	
	 The characterization  of this model is  relevant from an experimenter's perspective.
	 Indeed, by Definition \ref{DEF:minimal_attention} it is possible to retrieve DM's attention from data with small computational effort, by collecting for each menu all the items whose removal causes a switch.
	 Moreover, differently from the models of limited consideration listed in Section~\ref{SECT:Limited_consideration}, for choices with minimal limited attention the subject's consideration sets inferred from data are \textsl{unique} by definition, and all the attention filters that may overestimate the DM's attention are excluded.
\citet[Lemma~1 and Theorem~3]{MasatliogluNakajimaOzbay2012} characterize limited attention by the asymmetry and the acyclicity of the relation $P$ defined by
\begin{equation} \label{EQ:P_in_CLA} 
	x P y \quad \Longleftrightarrow \quad \text{ there is $A \in \X$ such that $x=c(A) \neq c(A \setminus y)$}
\end{equation}

The characterization of choices with minimal limited attention needs two additional conditions, displayed in the next result.

	\begin{theorem}\label{THM:CMLA_characterization}
		A choice function $c\colon\X\to X$ is a choice with minimal limited attention if and only if the following properties hold:
		\begin{itemize}
			\item [\rm(i)]  $P$ is asymmetric and acyclic;
			\item [\rm(ii)] there are no $A\in\X$, $x \in A\setminus c(A)$, and $y \in A$ such that $(A\setminus x,A)$ is not a switch, $(A\setminus xy,A\setminus x)$ is not a switch, and $(A\setminus y,A)$ is a switch;
			\item [\rm(iii)] there are no $A \in\X$, $x \in A\setminus c(A)$, and $y \in A$ such that $(A\setminus x,A)$ is not a switch, $(A\setminus y,A)$ is not a switch, and $(A\setminus xy,A\setminus x)$ is a switch.
		\end{itemize}
	\end{theorem}
	
Theorem~\ref{THM:CMLA_characterization}	shows that under conditions (i)--(iii) the DM's attention is minimal.
Condition (i) is already needed for the general model of limited attention.
If the removal of an alternative $y$ causes a switch from a menu in which $x$ is selected, then there is no menu in which $y$ is selected, and the removal of $x$ causes a switch.
Moreover, preference revealed from data cannot exhibit any cycle.
Condition (ii) states that if a minimal violation of Axiom $\alpha$ for $c$ occurs  removing an item $y$ from a menu $A$, then the same must happen removing an item $x\neq c(A)$ from $A$, or removing $y$ from $A\setminus x$.
Condition (iii) is dual: if drawing $y$ out of $A\setminus x$ causes a switch, then $(A\setminus x, A)$ or $(A\setminus y,A)$ must be a switch.

Once the experimenter checks that $c$ is a choice with minimal limited attention, she can uniquely recover the subject's attention $\Gamma^{\,min}$ simply setting $\Gamma^{\,min}(A)=S_A\cup c(A)$ for each menu $A$.

\begin{example}
Let $c\colon\X\to X$ be the choice defined on $X=\{x,y,z\}$ by $$\underline{x}yz,\;\;x\underline{y},\;\;x\underline{z},\;\;\underline{y}z.\footnote{The underlined items are those selected by the DM.
We omit menu containing only one item for brevity.}$$ 
The revealed preference $P$, defined by $xPy$ and $xPz$, is asymmetric and acyclic (as for any other choice on three items).
The pairs $(xy,xyz)$, and $(xz,xyz)$ are switches, thus conditions (ii) and (iii) of Theorem~\ref{THM:CMLA_characterization} hold.
We conclude that $c$ is a choice with minimal limited attention, explained by the pair $(\rhd,\Gamma^{\,\min})$,  consisting of any preference $\rhd$ that linearly extends $P$, and a minimal attention filter $\Gamma^{\,min}\colon\X\to \X$ uniquely defined by $$\Gamma^{\,min}(xyz)=S_{xyz}\,\cup \,c(xyz)=xyz,\;\Gamma^{\,min}(xy)=S_{xy}\,\cup\,c(xy)=y,\;\Gamma^{\,min}(xz)=S_{xz}\,\cup\,c(xz)=z,\;$$   $$\Gamma^{\,min}(yz)=S_{yz}\cup c(yz)=y.$$

According to the general model of limited attention $c$ can be explained also by the pair $(\rhd,\Gamma)$, consisting of a preference $\rhd$ on $X$ that linearly extends $P$, and the attention filter $\Gamma\colon\X\to\X$ defined by $\Gamma(A)=\Gamma^{\,min}(A)$ if $A\neq yz$, and $\Gamma(yz)=yz$.  
\end{example}
  
 In some situations, minimal limited attention does not fit data, and the observed choice is explained only by larger consideration sets, as showed in the next example.

 \begin{example}
Let $c\colon\X\to X$ be the choice defined on $X=\{x,y,z\}$ by $$\underline{x}yz,\;\;x\underline{y},\;\;\underline{x}z,\;\;\underline{y}z,$$

As before, the revealed preference $P$, defined by $xPy$, is asymmetric and acyclic.
The the pair $(xy,xyz)$ is the unique switch in $c$, and condition (ii) of Theorem~\ref{THM:CMLA_characterization} fails.
Thus, $c$ is a not a choice with minimal limited attention.
Indeed, the map $\Gamma^{\,*}$ defined by $\Gamma^{\,*}(A)=c(A)\cup S_A$ for any $A\in\X$ that can be obtained from $c$ is not a minimal attention filter, since condition (ii)(a) of Definition~\ref{DEF:minimal_attention} is violated by $\Gamma^{\,*}(xyz)=xy\neq y=\Gamma^{\,*}(xy)$.
In this case the DM's consideration cannot be immediately recovered from switches and the selected items.
The experimenter needs to formulate additional assumptions about the DM's preference to obtain a (non minimal) attention filter for $c$.
\end{example} 

Individuals with minimal consideration can be analytically described also within the framework analyzed by \cite{CherepanovFeddersenSandroni2013}.
We devote the next subsection to this analysis.

\subsection{Minimal rationalization theory}

According to the basic rationalization theory  of \cite{CherepanovFeddersenSandroni2013}, defined by property iv) listed in Section~\ref{SECT:Limited_consideration}, the DM picks in each menu the maximal  alternative among those that are admitted by some psychological constraint.
In their representation, the DM's preference is an asymmetric binary relation $\succ$ on the ground set, and her psychological constraint is a choice correspondence $\psi$ satisfying Axiom $\alpha$.
When the DM's preference is a linear order $\rhd$, her behavior is consistent with the so-called ordered rationalization theory, formally reported by property v) of Section~\ref{SECT:Limited_consideration}.
In what follows, we describe a DM whose psychological constraint admits the least possible items.
We first need some preliminary notation.

\begin{definition}\label{DEF:minimal_psychological_constraint}
	Given a choice $c\colon\X\to X$ on $X$, and a menu $A\in\X$, denote by $D_{A}$  the set $\{x\in A\colon (\exists B\supseteq A)\,x=c(B)\}$, and let $\psi^{\,min}\colon\X\to \X$ be the choice correspondence such that $\psi^{\,min}(A)=D_A$ for any menu $A\in\X$. 
\end{definition} 

The set $D_A$ collects all the items of $A$ that have been selected in some greater $B$ menu including $A$.
For each menu $A$, $D_A$ is the smallest consideration set compatible with the choice behavior explained by basic and ordered rationalization theory.
Indeed, we have

\begin{lemma}\label{LEM:minimal_psychological_constraint}
Let $c\colon \X \to X$ be a choice function.
The choice correspondence $\psi^{\,min}$ satisfies Axiom $\alpha$.
Moreover, if $c \colon \X \to X$ is consistent with basic rationalization theory, and $(\psi,\succ)$ is an explanation by consideration of $c$, then $\psi^{\,min}(A) \sbs \psi(A)$ for any $A\in\X$.
	If $c \colon \X \to X$ is consistent with ordered rationalization theory, and $(\psi,\rhd)$ is an explanation by consideration of $c$, then $\psi^{\,min} (A)\sbs \psi(A)$ for any $A\in\X$.
\end{lemma}

We now introduce a special case of basic and ordered rationalization theory, in which the DM considers as few items as possible.

\begin{definition}\label{DEF:minimal_basic_rationalization_theory}
A choice $c\colon \X\to X$ is \textit{consistent with minimal basic rationalization theory} (\textit{MBR}) if there is  an asymmetric binary relation $\succ$ on $X$ such that  $c(A)=\max(\psi^{\,min}(A),\succ)$ for any menu $A$.
We call $\psi^{\,min}$ a \textit{minimal psychological constraint}.
%
\end{definition}

\begin{definition}\label{DEF:minimal_ordered_rationalization_theory}
	A choice $c\colon \X\to X$ is \textit{consistent with minimal ordered rationalization theory} (\textit{MOR}) if there is a linear order $\rhd$ on $X$ such that $c(A)=\max(\psi^{\,min}(A),\rhd)$ for any menu $A$.
	We call $\psi^{\,min}$ a \textit{minimal psychological constraint}. 
%
\end{definition}

Definitions~\ref{DEF:minimal_basic_rationalization_theory} and \ref{DEF:minimal_ordered_rationalization_theory} formalize the behavior of a DM who chooses according to \textit{strict} psychological constraints, which deem acceptable the least possible amount of alternatives.
As for the case of minimal attention, once the experimenter checks that choice data can be justified by minimal basic rationalization theory or minimal order rationalization theory, he can uniquely infer the DM's consideration set by collecting from each menu, the items which have been selected in any greater menu including the analyzed one.
Thus, he can retrieve the unobserved subset of considered items by setting $\psi^{\,min}(A)=D_A$ for each $A\in \X$.
We look for the necessary and sufficient conditions that characterize choices consistent with minimal basic and ordered rationalization theory.
It turns out that minimal consideration sets can be always retrieved from the models of \cite{CherepanovFeddersenSandroni2013}. 
Indeed, we have

\begin{theorem}\label{THM:equivalence_minimal_rationalization_theory}
	A choice $c$ is consistent with minimal basic rationalization theory if and only if it consistent with basic ordered rationalization theory.
	A choice $c$ is consistent with minimal  ordered rationalization theory if and only if it consistent with ordered rationalization theory.
\end{theorem} 

Theorem \ref{THM:equivalence_minimal_rationalization_theory} shows that datasets justified by rationalization theory can be always explained by means of unique collections of consideration sets, if the experimenter relies on the hypothesis  that the DM's consideration is as narrow as possible.
Unique consideration sets can be also retrieved from a further specification of basic and ordered rationalization theory, in which strict psychological constraints and limited attention jointly affect individual decisions.  

\subsection{Minimally attentive rationalization theory}

Minimal limited attention and rationalization theory explain choice data by assuming that the DM is moved by her bounded perception or her severe psychological restrictions.
However, these two phenomena may interact, and, especially when the DM is presented with many distinct options, determine her choice.
Hereafter we investigate the behavior of a DM whose psychological constraint is strict, but it still comprise the least amount of alternatives that her attention can capture.
We first define the DM's consideration sets.

\begin{definition}
	Given a choice $c\colon\X\to X$ on $X$ and a menu $A\in\X$, let $S_{A}^{\uparrow}$ be the set $\{x\in A\colon (\exists B\supseteq A)\,x\in S_B\}$ .
	Moreover, let $\psi_{*}^{\,min}$ the choice correspondence on $X$ such that $\psi_{*}^{\,min}(A)=S_A \cup D_A$, for any $A\in\X$.
 \end{definition}
 
In the set $S_{A}^{\uparrow}$ are collected all the items of the menu $A$, whose removal from any greater menu $B$ including $A$ determines a switch.
The choice correspondence $\psi_{*}^{min}$ picks from each menu the items that either caused a switch or have been chosen in some larger menu that includes the examined one.
Moreover, the information contained in $\psi^{min}_{*}$ yields two consequences.
First, $\psi^{min}_{*}$ is a psychological constraint, as defined in \cite{CherepanovFeddersenSandroni2013}.
Second, the correspondence $\psi^{min}_{*}$ is the smallest psychological constraint the includes the least possible number of items that can be observed by a DM with minimal limited attention.
Indeed we have

\begin{lemma}\label{LEM:minimal_psychological_constraint_including_minimal_attention_filter}
	Let $c\colon\X\to X$ be a choice function on $X$.
	The choice correspondence $\psi_{*}^{min}$ satisfies Axiom $\alpha$.
	Moreover, for any choice correspondence $\psi\colon \X\to\X $ on $X$ satisfying Axiom $\alpha$  and such that $\psi(A)\supseteq \Gamma^{\,min}(A)$, we have that $\psi(A)\supseteq \psi_{*}^{min}(A)\supseteq \Gamma^{\,min}(A)$.
\end{lemma}

We now formally define the behavior of a DM endowed with a strict psychological constraint, which still contains the minimal amount of items that can be perceived trough her attention.

\begin{definition}\label{DEF:minimally_attentive_basic_rationalization_theory}
A choice $c\colon \X\to X$ is \textit{consistent with minimally attentive basic rationalization theory} (\textit{MABR}) if there is an asymmetric relation $\succ$ on $X$ such that  $c(A)=\max(\psi^{\,min}_{*}(A),\succ)$ for any menu $A$.
We call $\psi^{\,min}_{*}$ a \textit{minimally attentive psychological constraint}.
%
\end{definition}

\begin{definition}\label{DEF:minimally_attentive_ordered_rationalization_theory}
	A choice $c\colon \X\to X$ is \textit{consistent with minimally attentive ordered rationalization theory} (\textit{MAOR}) if there is a linear order $\rhd$ on $X$ such that $c(A)=\max(\psi^{\,min}_{*}(A),\rhd)$ for any menu $A$.
\end{definition}

Definitions \ref{DEF:minimally_attentive_basic_rationalization_theory} and \ref{DEF:minimally_attentive_ordered_rationalization_theory} describe a specification of basic and ordered rationalization theory in which the DM's psychological constraint is the smallest one including her minimal attention filter.
Choice data  justified by one of these models exhibit the effect of the interaction between limited attention and psychological restrictions on individual decisions.
Moreover, consideration sets  obtained by minimally attentive basic and ordered rationalization theory are uniquely identified.
Indeed, once he checks that a choice function satisfies Definition~\ref{DEF:minimally_attentive_basic_rationalization_theory} or ~\ref{DEF:minimally_attentive_ordered_rationalization_theory}, the experimenter can recover the DM's minimally attentive psychological constraint $\psi^{\,min}_{*}$ by setting $\psi^{\,min}_{*}(A)=S_A^{\uparrow}\cup D_A$ for each feasible menu $A$.
The characterization of minimally attentive basic and ordered rationalization theory relies on some properties of the following revealed preference relation. 

\begin{definition}\label{DEF:revealed_preference_minimally_attentive_rationalization theory}
	Let $c\colon\X\to X $ be a choice on $X$.
	Let $R$ be the binary relation on $X$ defined, for any $x,y\in X$, by $xRy$ if there are two menus $A,B\in \X$ such that $B\supseteq A$, $x,y\in A$, $x=c(A)$, and either $y=c(B)$ or $y\in S_{B}$.
\end{definition}

The binary relation $R$ gathers all the information expressed by the observed choice: an item $x$ is preferred to $y$ if they both belong to some menu from which $x$ is selected, and there is a greater menu in which either $y$ is selected or its removal causes a switch.  
The examination of the binary relation $R$ allows the experimenter to verify whether choice data can be explained by the models of limited consideration introduced in Definitions~\ref{DEF:minimally_attentive_basic_rationalization_theory} and~\ref{DEF:minimally_attentive_ordered_rationalization_theory}.
Indeed, we have:

\begin{theorem}\label{THM:minimally_attentive_rationalization_theory_characterization}
	A choice function $c$ is consistent with minimally attentive basic rationalization theory if and only if  $R$ is asymmetric.
	 A choice function $c$ is consistent with minimally attentive ordered rationalization theory if and only if $R$ is asymmetric and acyclic.
\end{theorem}  
   
 Minimal limited attention, minimal rationalization theory, and minimal attentive rationalization theory are models of limited consideration that enable the experimenter to estimate unique consideration sets from the observed choice with low computational effort, as resumed in Table~\ref{TABLE:summary_unique_consideration_sets}.

 \begin{table}[h]
\begin{center}
\begin{tabular}{|c|c|c|}
\cline{1-2}
$\mathscr{P}$ &  $\mathscr{F}$   \\
\hline
 CMLA &   $\Gamma^{\,min}$ & $\Gamma^{\,min}(A)=c(A)\cup S_A$\\
\hline
MBRT, MORT&  $\psi^{\,min}$ & $\psi^{\,min}(A)=D_A$\\
\hline
MABRT, MAORT   & $\psi^{\,min}_{*}$& $\psi^{\,min}_{*}(A)=D_A\cup S^{\uparrow}_A$\\
\hline
\end{tabular}
\end{center}
\caption{Each model of limited consideration $\mathscr{P}$ in the first column is associated to the respective family of consideration filters $\mathscr{F}$, displayed in the second column. 
In the third column, for each model we show how to infer, given an observed choice, the unique subset of considered items from a generic menu $A$.}
\label{TABLE:summary_unique_consideration_sets}
\end{table}

\clearpage

\newpage

\setcounter{page}{1}

\section*{Appendix: proofs}

\noindent{\underline{\it Proof of Lemma \ref{LEMMA:necessary_conditions_of_RLC}.}
Assume $c\colon\X\to X$ is with rational limited consideration and $(\Gamma^{\,\mathfrak{c}},\rhd)$ it is some explanation by consideration of $c$.
Moreover, let $A\in\X$ and $x\in A$ be such that $x\neq c(A)\neq c(A\setminus x)$.
Since $\Gamma^{\mathfrak{c}}$ satisfies axiom $\alpha$, we conclude that $c(A)\in\Gamma^{\,\mathfrak{c}}(A\setminus x)$.
Since $c$ is RLC and $c(A)\in\Gamma^{\,\mathfrak{c}}(A\setminus x)$, we must have $c(A\setminus x)\rhd c(A)$, and so $c(A\setminus x) \notin \Gamma^{\,\mathfrak{c}}(A)$.
Toward a contradiction, suppose $\Gamma^{\, \mathfrak{c}}(A)=c(A)$.
By hypothesis\, $\Gamma^{\, \mathfrak{c}}$ satisfies $\alpha, \gamma,$ and $\delta$.
Therefore there is a partial order $P$ such that $c(A)=\Gamma^{\, \mathfrak{c}}(A)=\max(A,P)$.
It is easy to show that if $\max(A,P)=c(A)$, $P$ is a partial order, and $A$ is finite, then $c(A) P y$ for all $y \in A\setminus c(A)$.
We obtain that $c(A) P c(A\setminus x)$, hence $c(A \setminus x) \notin \Gamma^{\,\mathfrak{c}}(A \setminus x)$, a contradiction.
\qed

\medskip
\noindent\underline{\it Proof of Lemma~\ref{LEM:CSLA_equivalent_representation}.}
The equivalence between (i) and (iii) has been proved in \citet[Theorem~3]{GiarlottaPetraliaWatson2022b}.
This result also contains a third equivalent condition, i.e., the asymmetry of the binary relation $\vDash$, defined by $x\vDash y$ if there is a menu $A\in\X$ such that $x,y\in A$ and $(A\setminus x, A)$ is a minimal switch.   
Note also that, according to \citet[Proposition~1]{RavidStevenson2021}, a choice function $c\colon\X\to X$  admits a general temptation representation if and only if  the Axiom of Revealed Temptation (ART) holds.
ART requires that for any $B\in\X$, there is $x\in B$ such that  any pair $(B^{\prime},B^{\prime\prime})$ with $x \in B^{\prime}\subset B^{\prime\prime}\subseteq B$ is not a switch.
Thus, to complete the proof of Lemma~\ref{LEM:CSLA_equivalent_representation} we only need to show that the asymmetry of $\vDash$ is equivalent to ART.

Let $c\colon\X\to X$ be a choice function.
We first show that when $\vDash$ is asymmetric, ART holds.
Assume that $\vDash$ is asymmetric.
By \citet[Lemma 4]{GiarlottaPetraliaWatson2022b}, we obtain that $\vDash$ is acyclic, hence a strict partial order.}
By \cite{Szpilrajn1930}'s theorem,  there is a linear order $\rhd$ that extends $\vDash$.
We claim that, for any $B\in \X$, any pair  $(B^{\prime},B^{\prime\prime})$ such that $\max(B,\rhd)\in B^{\prime}\subset B^{\prime\prime}\subseteq B$ is not a switch. 
Toward a contradiction, assume there are $B^{\prime}\subset B^{\prime\prime}\subseteq B$ such that $\max(B,\rhd)\in B^{\prime}$ and $(B^{\prime}, B^{\prime\prime})$ is a switch. 
By Lemma~\ref{LEMMA:minimal_violations_of_alpha}, there are $y \in X$ and $C \in \X$ such that $B^{\prime}\subseteq C \setminus y \subsetneq C \subseteq B^{\prime\prime}$ and $(C\setminus y,C)$ is a minimal switch. 
Note that $y$ must be distinct from $\max(B,\rhd)$.
The definition of $\vDash$ implies that $y \vDash x$, and that $y\rhd x$, a contradiction. 

To prove that ART implies the asymmetry of $\vDash$, assume toward a contradiction that ART holds, and $\vDash$ is not asymmetric.
Thus, there are $x,y\in X$, and $D,E\in \X$ such that $x,y\in D\cap E$, $(D\setminus x,D)$ is a minimal switch, and $(E\setminus y,E)$ is a minimal switch.
Consider the menu $D\cup E$.
Axiom $\alpha$ does not hold for the collection $\{G \subseteq D\cup E \colon x\in G\}$, since $(E\setminus y, E)$ is a minimal switch.
Axiom $\alpha$ does not hold for the collection $\{G \subseteq D\cup E \colon y\in G\}$, since $(D\setminus x, D)$ is a minimal switch.
Moreover, for any $z\in D\setminus \{x,y\}$ Axiom $\alpha$ fails for the collection $\{G \subseteq D\cup E \colon z\in G\}$, since $(D\setminus x, D)$ is a minimal switch.
Finally for any $z\in E\setminus \{x,y\}$ Axiom $\alpha$ fails for the collection $\{G \subseteq D\cup E \colon z\in G\}$, since $(E\setminus y, E)$ is a minimal switch.
We conclude that ART fails for $c$, a contradiction.
\qed

%

\medskip
\noindent\underline{\it Proof of Lemma~\ref{LEM:CSSLA_alpha}.}
$(\Longrightarrow)$.	 Suppose $c\colon\X\to X$ is a choice with selective salient attention, and let $\left(\Gamma^{\,sel},\rhd\right)$ be some explanation by consideration of $c$.
Assume toward a contradiction that axiom $\alpha$ fails for $c$.
Thus, by By Lemma \ref{LEMMA:minimal_violations_of_alpha} there are $A\in\X$ and $y\in  A$ such that $y\neq c(A)=x\neq c(A\setminus y)$. 
We conclude that $\Gamma^{\,sel}(A\setminus y)\neq\Gamma^{\,sel}(A)\setminus y$, which is false, since $x\neq y$.

$(\Longleftarrow)$. Assume $c\colon\X\to X$ satisfies axiom $\alpha$.
According to \cite{Samuelson1938}, there is a linear order $\rhd$ on $X$ that rationalizes $c$.
Let $\Gamma^{*}:\X\to X$ be a choice correspondence such that $\Gamma^{*}(A)=A$, for any $A\in \X$.
We have that $\Gamma^{*}$ is a selective salient attention filter, $(\Gamma^{*},\rhd)$ is an explanation by consideration of $c$, and, as a consequence, that $c$ is with selective salient attention.
\qed

\medskip
\noindent{\underline{\it Proof of Lemma \ref{LEM:CSSLA_non_trivial_consideration_sets}.}
Let $c\colon \X$ be a choice with selective salient limited attention.
By Lemma~\ref{LEM:CSSLA_alpha}, $c$ satisfies Axiom $\alpha$, and it is rationalizable.
Let $\succ$ be the linear order such that $c(A)=\max(A,\rhd)$ for each $A\in\X$.
Let $x_{\,min}=\min(X,\rhd)$ be the minimum element of $X$ with respect to $\rhd$.
Let $\Gamma^*\colon\X\to \X$ be a choice correspondence such that $\Gamma^*(A)=A\setminus x_{\,\min}$, for any $A\in\X\setminus \{x_{\,\min}\}$, and $\Gamma^*(x_{\,\min})=x_{\,\min}$.

We need to show that $\Gamma^*$ is a selective attention filter, and that $(\Gamma^*,\rhd)$ is an explanation by limited consideration of $c$.
To show that $\Gamma^*$ is a selective attention filter, take a menu $A\in\X$, and an item $x\in A$.
If $x=x_{\,\min}$, the definition of $\Gamma^*$ implies that $\Gamma^*(A)\setminus x=A\setminus{x}=\Gamma(A\setminus x)$.
If $x\neq x_{\,\min}$,  the definition of $\Gamma^*$ yields $\Gamma^*(A)\setminus x=A\setminus \{x_{\,\min},x\}=\Gamma^*(A\setminus x)$.
To show that $(\Gamma^*,\rhd)$ is an explanation by consideration of $c$, note that  $x_{\,\min}\neq c(A)=\max(A,\rhd)=\max(\Gamma^*(A),\rhd)$ for any $A\in\X\setminus \{x_{\,\min}\}$, and $c(x_{\,\min})=c(\Gamma^*(x_{\,\min}),\rhd)=x_{\,\min}$.
\qed

\medskip
\noindent{\underline{\it Proof of Theorem \ref{THM:list_rationalizable}.}
We first need some preliminary notions and results.
    
    \begin{definition}
Assume $c\colon\X\to X$ is with competitive limited attention, and let $(\Gamma^{{\,co}},\rhd)$ be some explanation by consideration of $c$.
We say that $\Gamma^{\,co}$ is a \textsl{maximal competitive attention filter} for $c$ if there is no competitive attention filter $\Gamma^{{{\,co}}^{\prime}}\colon\X\to\X$ distinct from $\Gamma^{{\,co}}$ such that $(\Gamma^{{{\,co}}^{\prime}},\rhd)$ is an explanation by consideration of $c$, and $\Gamma^{{\,co}}(A)\subseteq \Gamma^{{{\,co}}^{\prime}}(A)$ for all $A \in \X$.\end{definition}

An analogue definition holds for choice with limited attention.

\begin{definition}
Assume $c\colon\X\to X$ is a choice with limited attention and $(\Gamma,\rhd)$ is some explanation by consideration of $c$.
We say that $\Gamma$ is a \textsl{maximal attention filter} for $c$ if there is no attention filter $\Gamma^{\prime}\colon\X\to\X$ distinct from $\Gamma$ such that $(\Gamma^{\prime},\rhd)$ is an explanation by consideration of $c$ and $\Gamma(A)\subseteq \Gamma^{\prime}(A)$ for all $A \in \X$.\end{definition}

Thus, a maximal (competitive) attention filter describes the largest extent of DM's attention compatible with choice data.  
Given a linear order $\rhd$ on $X$, we denote by $x^\downarrow$ the set $\{y \in X : x\, \triangleright\, y\} \cup \{x\}$, and  let $\Gamma_\rhd (A) = c(A)^\downarrow \cap A$. 
For choices with limited attention,  \citet[Lemma 17]{GiarlottaPetraliaWatson2022b} prove that $\Gamma_\rhd$ is a maximal attention filter.


\begin{lemma}\label{LEMMA:CLA_extension}
	Assume $c\colon \X\to X$ is a choice with limited attention, and let $P$ be the relation defined by $x P y$ if there is $A \in \X$ such that $x=c(A) \neq c(A \setminus y)$.
	 For any  linear order $\triangleright$ that extends $P$ we have that $\Gamma_\rhd$ is a maximal attention filter for $c$, and $(\Gamma_{\rhd},\rhd)$ is an explanation by consideration of $c$.
\end{lemma} 

A result analogue to Lemma 	\ref{LEMMA:CLA_extension} holds for choices with competitive limited attention.

\begin{lemma}\label{LEMMA:LR_maximal_filter}
	If $c$ is a choice with competitive limited attention , and $(\triangleright, \Gamma^{\,co})$ is some explanation by consideration of $c$, then $\Gamma_{\triangleright}$ is a maximal competitive attention filter for $c$, and $(\Gamma_{\triangleright},\rhd)$ is an explanation by consideration of $c$.
\end{lemma}

\begin{proof}
	
We need to show that
	 (i) $c(A)=\max\left(\Gamma_{\rhd}(A),A\right)$ for any $A\in\X$, 
	(ii) $\Gamma_{\triangleright}$ is a competitive attention filter, and
	 (iii) $\Gamma_{\triangleright}$ is maximal for $c$.
	
\begin{enumerate}[\rm(i)]

\item The proof readily follows from the definition of $\Gamma_\rhd$.

\item We need to show that properties (a), (b), and (c) listed in Definition \ref{DEF:list_attention} hold for $\Gamma_{\rhd}$.
\begin{itemize}

\item[(a)] Let $A \in \X$ be any menu, and $x$ an item of $A$ different from $\min(A,\rhd)$.
Toward a contradiction, suppose $\Gamma_\rhd(A)\setminus x \neq\Gamma_\rhd(A\setminus x)$.
The definition of $\Gamma_\rhd$ yields $ c(A \setminus x)^{\downarrow}\cap (A\setminus x)\neq \left(c(A)^{\downarrow}\cap A\right)\setminus x$, hence $c(A)\neq c(A\setminus x)$.
Moreover, we have $x\neq \max(\Gamma(A),\rhd)$. 
Since $\Gamma^{\,co}(A)\setminus x=\Gamma^{\,co}(A \setminus x)$ because $\Gamma^{\,co}$ is a competitive attention filter, we obtain $c(A)\in\Gamma^{\,co}(A\setminus x)$ and $c(A\setminus x)\in\Gamma^{\,co}(A)$, which respectively yield $c(A\setminus x)\rhd c(A)$ and $c(A)\rhd c(A\setminus x)$, a contradiction. 
\item[(b)] Assume toward a contradiction that there is $A\in\X$ such that $\Gamma_\rhd(A)\neq \min(A,\rhd)$ and $\Gamma_\rhd(A\setminus\min(A,\rhd))\neq \Gamma_\rhd(A)\setminus\min(A,\rhd)$.
Thus, there is $x\in A$ distinct from $\min(A,\rhd)$ such that $c(A)\rhd x$, and $c(A \setminus\min(A,\rhd))^{\downarrow}\cap (A\setminus \min(A,\rhd))\neq \left(c(A)^{\downarrow}\cap A\right)\setminus \min(A,\rhd)$, which yields $c(A)\neq c(A\setminus\min(A,\rhd))$. 
Since $c$ is CCLA, either (1) $x=\Gamma^{\,co}(A)=\min(A,\rhd)$ or (2) $\Gamma^{\,co}(A\setminus\min(A,\rhd))=\Gamma^{\,co}(A)\setminus\min(A,\rhd)$.
If (1) holds, we conclude that $c(A)=\min(A,\rhd)$, which is impossible, since we know that $c(A)\rhd x$.
If (2) holds, the definition of $\Gamma^{\,co}$ yields  $\min(A,\rhd)\neq c(A)\rhd c(A\setminus\min(A,\rhd))$ and $c(A\setminus\min(A,\rhd))\rhd c(A)$, which is impossible.
\item[(c)] Assume that $y\rhd x$ and $y\not\in \Gamma_\rhd(xy)$ (thus $c(xy)=x$).
Assume that $A\subseteq\,^{\rhd}]x,y[$.
Two cases are possible: (1) $c(A\cup xy)=y$, or (2) $c(A\cup xy)\neq y$.
Assume (1) holds.
Since $c$ is a choice with competitive limited attention, either (1)$^\prime$ $\Gamma^{\,co}(xy)= x$ or (1)$^{\prime\prime}$ $\Gamma^{\,co}(xy)\neq x$.
If (1)$^\prime$ is true, since $\Gamma^{\,co}$ is a competitive limited attention filter, we have that $y\not\in\Gamma^{\,co}(A\cup xy)$, and $c(A)\neq y$, which contradicts $(1)$.
If (1)$^{\prime\prime}$ holds, since $y\rhd x$, we have that $c(xy)=y$, which is false.
If (2) holds, we must have that $y\rhd c(A\cup xy)$.
The definition of $\Gamma_{\rhd}$ yields $y\not\in\Gamma_{\rhd}(A\cup xy)$. 

\end{itemize}
\item Suppose by way of contradiction that there is a competitive attention filter $\Gamma^{\,co^\prime}$ for $c$ such that $(\Gamma^{\,co^\prime},\rhd)$ is an   explanation by consideration of $c$ and $y\in \Gamma^{\,co^\prime}(A)\setminus \Gamma_\rhd(A)$ for some $A \in \X$ and $y\in A$.
Since $y\notin \Gamma_\rhd(A)$, we get $y\rhd c(A)$.
On the other hand, since $y \in \Gamma^{{\,co}^\prime}(A)$, $c$ is a choice with competitive limited attention, and $(\Gamma^{\,co^\prime},\rhd)$ is an explanation by consideration of $c$, we must have $c(A)\rhd y$ or $c(A) = y$, which is impossible.
\end{enumerate}
\end{proof} 

The next result holds for maximal attention filters associated to choices with competitive limited attention.

\begin{lemma}\label{LEMMA:LR(LA)_property}
	Assume $c\colon\X\to X$  is a choice with competitive limited attention and $(\Gamma^{\,co},\rhd)$ is some explanation by consideration of $c$.
	 If there is $A\in \X$ such that $c(A)\neq \min(A,\rhd)$, then $\Gamma_\triangleright(A\setminus\min(A,\rhd))=\Gamma_\triangleright(A)\setminus\min(A,\rhd)$.
\end{lemma}

\begin{proof}
Towards a contradiction, assume $c\colon\X\to X$ is a choice with competitive limited attention, and there is $A\in\X$ such that $c(A)\neq \min(A,\rhd)$ and $\Gamma_\triangleright(A\setminus\min(A,\rhd))\neq\Gamma_\triangleright(A)\setminus\min(A,\rhd)$.
	By Lemma \ref{LEMMA:LR_maximal_filter},  we have that $\Gamma_\triangleright(A)=\min(A,\rhd)$.
	Since $c(A)\neq \min(A,\rhd)$ and $c(A) \in \Gamma_\triangleright(A)$, we have a contradiction.
\end{proof}


Thus, the worst item never affects DM's attention, even when we consider the widest range of observed alternatives. 
An immediate corollary of Lemma \ref{LEMMA:LR(LA)_property} is

\begin{corollary}\label{COR:LR(LA)_property}
	Assume $c$ is a choice with competitive limited attention, and let $(\Gamma^{\,co},\rhd)$ be some explanation by consideration of $c$.
	For any $A\in \X$, and $x \in A\setminus c(A)$, we have that
	$$
	\Gamma_{\rhd}(A\setminus(x^{\downarrow} \cap A))=\Gamma_{\rhd}(A)\setminus(x^{\downarrow} \cap A).
	$$
\end{corollary}

\begin{proof}
Set $x_1=x$ and let $x^\downarrow \cap A=\{x_1,\dots,x_i,  \dots ,x_n\}$ be such that $x_{i} \triangleright x_{i+1}$ for any $i\in~\{1,\dots,n\}$.
	Apply Lemma~\ref{LEMMA:LR(LA)_property} to $x_n$ to obtain $\Gamma_\triangleright(A\setminus x_n)=\Gamma_\triangleright(A)\setminus x_n$.
	Next, apply Lemma~\ref{LEMMA:LR(LA)_property} to the set $A\setminus x_n$ and $x_{n-1}$ to obtain $\Gamma_\triangleright(A\setminus x_{n-1}x_n)=\Gamma_\triangleright(A)\setminus x_{n-1}x_n$.
	Iterating the process we obtain what we are after
	\end{proof}
	\smallskip


For maximal filters of choices with competitive limited attention, removing items dominated in the DM's judgement does not affect her attention.    
%
The next lemma shows that maximal attention filters satisfy a specific regularity property.
\begin{lemma}\label{LEMMA:CLA_property}
	
	If $c$ is a choice with limited attention and $(\Gamma,\rhd)$ is some explanation by consideration of $c$, 
	then for all $A \in \X$ we have $\Gamma_{\rhd}(A)=\Gamma_{\rhd}(\Gamma_{\rhd}(A))$.
\end{lemma}

\begin{proof}

Denote by $A_{*}$  the set
$$
A\setminus\{x_1,\cdots,x_i,\cdots, x_n\colon x_i \in (A\setminus\Gamma_\triangleright(A))\,(\forall\, 1\leq i\leq n)\}.
$$
Since $c$ is a choice with limited attention, and by Lemma \ref{LEMMA:CLA_extension} $\Gamma_{\rhd}$ is an attention filter for $c$, we have that $\Gamma_\triangleright (A)=\Gamma_\triangleright (A\setminus x_1)$.
Since $\Gamma_\triangleright (A)=\Gamma_\triangleright (A\setminus x_1)$, and $x_2\not \in \Gamma_\triangleright(A)$, we apply again the definition of attention filter to conclude that $\Gamma_\triangleright(A)=\Gamma_\triangleright (A\setminus x_1)=\Gamma_{\triangleright}(A\setminus x_1x_2)$.
We can repeat this argument until we get $\Gamma_{\triangleright}(A)=\cdots=\Gamma_{\triangleright}(A\setminus x_1\cdots x_{n})=\Gamma
_\triangleright(A\setminus A_{*})=\Gamma_{\triangleright}(\Gamma_\triangleright(A))$. 
\end{proof}
\medskip

List-rational choice functions are characterized by the absence of cycles in the binary relation \textsl{revealed to follow}, described below.

\begin{definition}[\citealp{Yildiz2016}]\label{DEF:revealed_to_follow}
   We say that $xFy$ (\textsl{x is revealed to follow to y}) if and only if at least one of the three following properties hold:
    \begin{itemize}
    	\item[\rm(i)] there exists $A \in \X$, such that $x=c(A \cup y)$ and $y=c(xy)$;
    	\item[\rm(ii)] there exists $A\in\X$, such that $x=c(A \cup y)$ and $x \neq c(A)$;
    	\item[\rm(iii)] there exists $A \in\X$, such that $x\neq c(A \cup y)$, $x=c(xy)$, and $x=c(A)$.
    \end{itemize}
\end{definition}

\citet[Corollary~1]{Yildiz2016} shows that LR choice functions are characterized by the asymmetry and acyclicity of $F$.
The next lemma clarifies the connection between list rationality and choices with competitive limited attention.

\begin{lemma}\label{LEMMA:LR_property}
	Assume $c$ is LR by some $\mathbf{f}$ list on $X$. For any $x,y\in X$
	we have that 
	$$
	c(xy)=x \implies (\forall A \subseteq\,^\mathbf{f}]x,y[) \; y\neq c(A \cup xy).
	$$
\end{lemma}

\begin{proof}
By LR, we have $c(A)=c(\{x,c(A\setminus x)\})$.
If $c(A \setminus x)\neq y$, then we are done.
Otherwise $c(A)=c\big(\{c(A \setminus x),\,x\}\big)=c(xy)=x$.
\end{proof}

The next result is taken from \citet[Corollary 1]{Yildiz2016}.

\begin{lemma}\label{LEMMA:LR_extension}
	If $c$ is LR for some list $\mathbf{f}$, then   $F \subseteq \mathbf{f}$.
\end{lemma}

The following result links choices with limited attention to list rationality.

\begin{lemma} \label{LEMMA: P_in_F}
	$P \subseteq F$.
\end{lemma}


\begin{proof}
	
	If $xPy$ holds for some $x,y\in X$, then there is $A\in\X$ such that $x=c(A)$ and $x \neq c(A \setminus y)$.
	Let $B=A\setminus y$ and observe that $x=c(B \cup y)$ while $x\neq c(A)$.
	Hence, $xFy$.
	\end{proof}

Finally we conclude:
    
\begin{corollary} \label{COR:LR_implies_CLA}
	If a choice is list rational, the it is a choice with limited attention.
\end{corollary}
 \begin{proof}
	Apply \citet[Corollary~1]{Yildiz2016} to obtain that $F$ is acyclic and antisymmetric.
	By Lemma~\ref{LEMMA: P_in_F}, $P$ is contained in $F$.
	Hence, $P$ is acyclic and antisymmetric, and  \citet[Lemma~1 and Theorem~3]{MasatliogluNakajimaOzbay2012} yields that $c$ is a choice with limited attention.
	 \end{proof}

We are now ready to prove the equivalence between list rationality and choices with competitive limited attention.

$(\Longrightarrow)$.
Assume $c\colon\X\to X$ is list rational for some list $\mathbf{f}$.    
    Thus, by Corollary~\ref{COR:LR_implies_CLA} $c$ is CLA.
    Lemma~\ref{LEMMA:LR_extension} and Lemma~\ref{LEMMA: P_in_F} yield $P\subseteq \mathbf{f}$.
   By Lemma~\ref{LEMMA:CLA_extension} $\Gamma_{\mathbf{f}}$ is a maximal attention filter for $c$, and  $\left(\Gamma_{\mathbf{f}},\mathbf{f}\right)$ is an explanation by consideration of $c$. 
To show that $c$ is with competitive limited attention, we only need to prove that properties (b) and (c) of Definition~\ref{DEF:list_attention} hold.

  \begin{itemize}
  	\item [(b)] The definition of list rationality implies that $c(A)=c(\{c(A\setminus \min(A,\mathbf{f})),\min(A,\mathbf{f})\})$.
    If $c(A)=c(A\setminus\min(A,\mathbf{f}))$, the definition of  $\Gamma_{\mathbf{f}}$ implies that $\Gamma_{\mathbf{f}}(A\setminus\min(A,\mathbf{f}))=\Gamma_{\mathbf{f}}(A)\setminus\min(A,\mathbf{f})$.
    If $c(A)=\min(A,\mathbf{f})$, the definition of $\Gamma_{\mathbf{f}}$ yields  that $\max\Gamma_{\mathbf{f}}(A)=\min(A,\mathbf{f})$.
    
   \item[(c)] Let $x,y \in X$ and $A \sbs\,^{\mathbf{f}}]x,y[$ such that $y \notin \Gamma_{\mathbf{f}}(x,y)$ (thus $c(A)=x$).
    We conclude that $y\,\mathbf{f}\,x$.
    Toward a contradiction, suppose $ y \in \Gamma_\mathbf{f}(A \cup xy)$.
    Since $A\sbs\,^{\mathbf{f}}]x,y[$, then $y= \max \Gamma_\mathbf{f}(A \cup xy)$.
    The definition of $\Gamma_{\mathbf{f}}$ yields $y=c(A \cup xy)$, a contradiction with Lemma~\ref{LEMMA:LR_property}.
    \end{itemize}
    
    $(\Longleftarrow)$.
    Suppose there $c$ is a choice with competitive limited attention, and  $(\Gamma^{\,co},\rhd)$ is some explanation by consideration of $c$.
    By Lemma~\ref{LEMMA:LR_maximal_filter}, $\Gamma_{\rhd}$ is a maximal competitive attention filter for $c$, and $(\Gamma_{\rhd},\rhd)$ is an explanation by consideration of $c$.
    We shall prove that $F \subseteq \triangleright$, and, by \citet[Corollary~1]{Yildiz2016}, $c$ is list rational.
    By contrapositive, suppose $\lnot(x \triangleright y)$ (thus $y \triangleright x$).
    Towards a contradiction, assume $xFy$.
    We prove that none of the three following properties are true:
    \begin{itemize}
    	\item[\rm(i)] there exists $A\in\X$ such that $x=c(A \cup y)$ and $y=c(xy)$;
    	\item[\rm(ii)] there exists $A\in \X$ such that $x=c(A \cup y)$ and $x \neq c(A)$;
    	\item[\rm(iii)] there exists $A\in \X$ such that $x\neq c(A\cup y)$, $x=c(xy)$, and $x=c(A)$.
    \end{itemize}
    
    \begin{enumerate}[\rm(i)]
    	\item Note that 
   by Lemma~\ref{LEMMA:CLA_property} we  can set without loss of generality $\Gamma_\triangleright(A \cup y)=\Gamma_\triangleright(A)=A$.\footnote{If $A\neq \Gamma_\triangleright(A)$, relabel $A$ as  $A \cap x ^\downarrow$ and observe that now $\Gamma_\triangleright(A)=A$ and $x=c(A \cup y)$.} 
    
    Denote by $x^{\downarrow}_{*}$ the set $\{y \in X : x\, \triangleright\, y\}$, and let $z=\max\big(A\cap\,x^{\downarrow}_{*},\rhd\big)$.
     Apply Corollary~\ref{COR:LR(LA)_property} to $z$ to obtain $\Gamma_\triangleright((A\,\cup\,y)\setminus (z^{\downarrow} \cap (A \cup\, y))=\Gamma_\triangleright(A\,\cup\,y)\setminus (z^{\downarrow} \cap (A\,\cup \,y))$.
    Since $x=\max(A,\rhd)$, we obtain $\Gamma_\triangleright(xy)= \Gamma_\triangleright((A \cup y)\setminus (z^{\downarrow} \cap (A \cup y))$.
    However, $y \in \Gamma_\triangleright(xy)$, while $y \notin \Gamma_\triangleright(A\,\cup\,y)\setminus (z^\downarrow \cap\,(A \cup\, y))$, a contradiction.
%
%
    
\item Assume toward a contradiction that there are $A\in\X$, $y\in X$, and $x\in A$ such that $x=c(A	\cup y)$ and $x\neq c(A)$.
    Since $c$ is a choice with competitive limited attention, and $y\not\in \Gamma_{\triangleright}(A\cup x)$, we must have that $\Gamma_{\triangleright}(A\cup y)=\Gamma_{\triangleright}(A)$, that yields $c(A\cup y)=c(A)$, which is false.
    
    \item Assume toward a contradiction that there are $A\in\X$, $y\in X$, and $x\in A$ such that $c(A\cup y)\neq x$, $c(A)=x$, and $c(xy)=x$.
     If $y \notin \Gamma_\triangleright(A \cup y)$, since $c$ is a choice with competitive limited attention we have that $\Gamma_\triangleright(A \cup y)=\Gamma_\triangleright(A)$, which is a contradiction, since $x=c(A)$ and $x \neq c(A \cup y)$.
    Thus, suppose $y \in \Gamma_\triangleright(A \cup y)$. 
     Let $z=\max\big(A\cap\,x^{\downarrow}_{*},\rhd\big)$.
     Apply Corollary~\ref{COR:LR(LA)_property} to $z$ and obtain $\Gamma_\triangleright\left((A\cup y)\setminus (z^{\downarrow} \cap (A \cup y)\right)=\Gamma_\triangleright(A\,\cup\,y)\setminus (z^{\downarrow} \cap (A \cup y))$.
     First, note that $y\in \Gamma_\triangleright(A\,\cup\,y)\setminus (z^{\downarrow} \cap (A \cup y))$.
     Second, observe that $\Gamma_\triangleright((A \cup y)\setminus (z^{\downarrow} \cap (A \cup y))=\Gamma_\triangleright((A\cap\,^{\rhd}]x,y[) \cup xy)$.
    Since $c$ is a choice with competitive limited attention, and $y \notin \Gamma_\triangleright(xy)$, property (ii)(c) of Definition~\ref{DEF:list_attention} yields $y\not\in\Gamma_\triangleright((A\,\cap\, ^{\triangleright}]x,y[) \cup xy)$, which is impossibile, since we already know that $y \in \Gamma_\triangleright(A \cup y)\setminus (z\downarrow \cap (A \cup y))=\Gamma_\triangleright((A\,\cap\, ^{\triangleright}]x,y[) \cup xy)$.
     
    Therefore, none of the three properties hold, and $\neg (xF y)$, which contradicts our hypothesis.
    We conclude that $\lnot(x \triangleright y)$ implies $\lnot(xFy)$.
    Hence, $F \subseteq \triangleright$, which implies that $F$ is asymmetric and acyclic.
    By \citet[Corollary~1]{Yildiz2016}, $c$ is list rational.\qed
     \end{enumerate}

\medskip
\noindent\underline{\it Proof of Lemma~\ref{LEMMA:necessary_conditions_of_CCLA}.}
Assume that $c\colon\X\to X$ is a choice with competitive limited attention, and let $A\in\X$ and distinct $x,y\in~A$ be such that $y=c(A)\neq c(A\setminus x)$.
The definition of choice with competitive limited attention implies that $\Gamma^{\,co}(A)\neq \Gamma^{\,co}(A\setminus x)$.
Property (a) and (b) of Definition \ref{DEF:list_attention} respectively yields $x\in\Gamma^{\,co}(A)$ and $x\neq \min(\Gamma^{\,co}(A),\rhd)$.
Since $x\in\Gamma^{\,co}(A)$ and $y=c(A)$, we conclude that $y\rhd x$.

Assume now that there are $x,y\in X$, and $A\in\X$ such that $y=c(xy)\neq c(A\cup xy)=x$.
The definition of choice with competitive limited attention implies that $x\in\Gamma^{\,co}(A\cup xy)$, and $y\in \Gamma^{\,co}(xy)$.
Two cases are possible: (i) $A\subseteq \mathring{(xy)}_{\rhd}$, or (ii) $A\setminus \mathring{(xy)}_{\rhd}\neq \es$.
If (i) holds, property $(c)$ of Definition~\ref{DEF:list_attention} implies that $x\in\Gamma^{\,co}(xy)$.
The definition of choice with competitive limited attention yields that $y\rhd x$.
Moreover, since $c(A\cup~xy)=x$, the definition of choice with competitive limited attention implies that $y\not\in\Gamma^{\,co}(A\cup xy)$. 
If (ii) is true, we are done.
\qed

\medskip
\noindent \underline{\it Proof of Theorem~\ref{THM:bounded_rationality_models_property_limited_consideration}.}
	Properties iii), iv), v), viii), ix) are models of limited consideration by definition.
	The equivalence between ii), iv), and xii) implies that ii) and xii) are models of limited consideration.	
Lemma \ref{LEM:CSSLA_alpha}, Theorem~\ref{THM:list_rationalizable}, Lemma~\ref{LEM:rational_attention_equivalent_shortlisting}, and Lemma~\ref{LEM:CSLA_equivalent_representation} show respectively that properties i), vi), vii), x), and xi) are models of limited consideration.
	\qed

\medskip
\noindent \underline{\it Proof of Lemma~\ref{LEM:minimal_attention}.}
Assume that $c\colon\X\to X$ is a choice with limited attention, and let $(\Gamma,\rhd)$ be an explanation by consideration of $c$.
Assume that $x\in S_{A}\cup c(A)$, for some $x\in X$, and $A\in\X$.
Thus, by Definition \ref{DEF:CLA_switch_sets}, we have that either $x=c(A)$ or $(A\setminus x,A)$ is a switch.
In the first case, since $(\Gamma,\rhd)$ is an explanation by consideration of $c$, we have that $x\in\Gamma(A)$.
If the second case holds, we have that $(A\setminus x,A)$ is a switch.
By property iii) listed in Section~\ref{SECT:Limited_consideration}, we have that $x\in\Gamma(A)$.  
\qed

\medskip
\noindent \underline{\it Proof of Theorem~\ref{THM:CMLA_characterization}.}
	$(\Longrightarrow)$. Suppose $c\colon \X\to X$ is a choice with minimal limited attention.
	Thus, there are a linear order $\rhd$ on $X$ and a minimal attention filter $\Gamma^{\,min}\colon\X\to\X$  such that $c(A)=\max(\Gamma^{\,min}(A),\rhd)$.
	We prove that conditions (i)--(iii) hold.
	\begin{description}
		\item [\rm(i)] Since $c$ is a choice with limited attention, by \citet[Lemma 1 and Theorem 3]{MasatliogluNakajimaOzbay2012} the binary relation $P$ is asymmetric and acyclic.
		\item [\rm(ii)] Toward a contradiction, let $A \in\X$, $x\in A \setminus c(A)$, $y \in A$ and suppose that $(A\setminus x,A)$ is not a switch, $(A\setminus xy,A\setminus x)$ is not a switch, and $(A\setminus y,A)$ is a switch.
		    The definition of $\Gamma^{\,min}$ yields that $x \notin \Gamma^{\,min}(A)$.
		    Since $y\in S_A$, we have that $y\in\Gamma^{\,min}(A)$.
		    If $c(A\setminus x)=y$, since $(A,A\setminus x)$ is not a switch, we would obtain $y=c(A)$, a contradiction with $y \in S_A$.
		    We conclude that $y\neq c(A\setminus x)$.
		    Moreover, since $(A\setminus xy,A\setminus x)$ is not a switch, we know that $y\not\in S_{A\setminus x}$.
		    The definition of $\Gamma^{\,min}$ implies that $y\not\in\Gamma^{\,min}(A \setminus x).$
		    
		     We conclude that $y \in \Gamma^{\,min}(A) \setminus \Gamma^{\,min}(A \setminus x)$, which is impossible, since $\Gamma^{\,min}$ is minimal attention filter.
		\item [\rm(iii)] Toward a contradiction, let $A\in\X$, $x\in A \setminus c(A)$, $y \in A$ and suppose that $(A\setminus x,A)$ is not a switch, $(A\setminus y,A)$ is not a switch, and $(A\setminus xy,A\setminus x)$ is a switch.
		    We obtain $x\notin \Gamma^{\,min}(A)$.
		    If $c(A)=y$,  since $(A\setminus x,A)$ is not a switch, we would have $y=c(A\setminus x)$, and $(A\setminus xy,A\setminus x)$ would not be a switch, which is a contradiction.
		    Since $y\not\in S_A$ and $y\neq c(A)$, we conclude that $y \notin \Gamma^{\,min}(A)$.
		    Since $(A\setminus xy,A\setminus x)$ is a switch, we have that $y\in \Gamma^{\,min}(A\setminus x)$.
		    We obtain that $y\in\Gamma^{\,min}(A \setminus x)\setminus\Gamma^{\,min}(A)$, which is impossible, since $\Gamma^{\,min}$ is a minimal attention filter.
%
	\end{description}
    $(\Longleftarrow)$. Suppose $c\colon\X\to X$ satisfies properties (i)--(iii) of Theorem \ref{THM:CMLA_characterization}.
    By assumption $P$ is asymmetric and acyclic.   
     Let $\rhd$ be a linear order that extends $P$.
    Note that for all $A \in\X$ we have $c(A)=\max(\Gamma^{\,min}(A),\rhd)$.
    
    To conclude the proof, we first show that for all $A \in\X$, if $x \notin \Gamma^{\,min}(A)$, then $\Gamma^{\,min}(A)=\Gamma^{\,min}(A\setminus x)$.
   Assume that $y \notin \Gamma^{\,min}(A)$.
   We want to prove that $y \notin \Gamma^{\,min}(A\setminus x)$.
    Since $x \notin \Gamma^{\,min}(A)$, we obtain that $(A\setminus x,A)$ is not a switch and $x \neq c(A)$.
    Using the same argument, $(A\setminus y,A)$ is not a switch and $y \neq c(A)$.
    Property (ii) implies that $(A\setminus xy,A\setminus x)$ is not a switch, hence either $y=c(A\setminus x)$ or $y \notin \Gamma^{\,min}(A\setminus x)$.
   If $y=c(A\setminus x)$, since $c(A) \neq x$ and $(A\setminus x,A)$ is not a switch, we must have that $c(A)=c(A\setminus x)=y$, which is a contradiction, since $y \notin \Gamma^{\,min}(A)$.
    
   Assume now that $y \notin \Gamma^{\,min}(A\setminus x)$.
    Since $x\not\in\Gamma^{\,min}(A)$, we know that $(A\setminus x,A)$ is not a switch.
    Since $y \notin \Gamma^{\,min}(A\setminus x)$, the pair $(A\setminus xy,A\setminus x)$ is not a switch.
    Property (i) yields $(A\setminus y,A)$ is not a switch.
    If $y \notin \Gamma^{\,min}(A)$, we are done.
    Toward a contradiction, suppose $y=c(A)$.
    Since $x \notin \Gamma^{\,min}(A)$, then $y=c(A)=c(A\setminus x)$, a contradiction with $y \notin \Gamma^{\,min}(A\setminus x)$.
    \qed

\medskip
\noindent \underline{\it Proof of Lemma~\ref{LEM:minimal_psychological_constraint}.}
Let $c\colon\X\to X$ be a choice function on $X$
To show that $\psi^{\,\min}$ satisfies Axiom $\alpha$, take $A,B\in\X$, and $x\in A$ such that $B\supseteq A$, and $x\in\psi^{\min}(B)$. 
  Definition~\ref{DEF:minimal_psychological_constraint} implies that there is $C\supseteq B$ such that $c(C)=x$.
  Since $C\supseteq A$, we apply again Definition~\ref{DEF:minimal_psychological_constraint} to conclude that $x\in\psi^{\,\min}(A)$.

Assume that $c$ is consistent with basic rationalization theory, and let $(\psi,\succ)$ be an explanation by consideration of $c$.
Assume that $x\in \psi^{\,\min}(A)$, for some $A\in\X$ and $x\in A$.
Definition~\ref{DEF:minimal_psychological_constraint} implies that there is some $B\supseteq A$ containing $x$ such that $c(B)=x$.
Since $\psi$ satisfies Axiom $\alpha$, we have that $x\in\psi(A)$.
The same argument can be used when $c$ is consistent with ordered rationalization theory.\qed
\medskip

\noindent \underline{\it Proof of Theorem~\ref{THM:equivalence_minimal_rationalization_theory}.}
We need some preliminary notions and result.
Given a choice function $c\colon\X\to X$ and  choice correspondence $\psi\colon\X\to \X$ defined on  a ground set $X$, denote by $\succ_{c,\psi}$ the binary relation defined, for any $x,y\in X$, by $x\succ_{c,\psi}y$ if there is $A\in X$ such that $c(A)=x$, and $y\in \psi(A)\setminus x$.
The following results holds.
\begin{lemma}\label{LEM:asymmetry_revealed_preference}
	Let $c\colon \X\to X$ and $\psi\colon \X \to \X$ be respectively a choice function and a choice correspondence on $X$.
	For any menu $A\in\X$, we have that $c(A)=\max(\psi(A),\succ_{c,\psi})$ if and only if $\succ_{c,\psi}$ is asymmetric. 
\end{lemma}  
\begin{proof}
Let $c\colon \X\to X$ and $\psi\colon \X \to \X$ be respectively a choice function and a choice correspondence on $X$.

($\Longrightarrow$).	
If $x\succ_{c,\psi} y$ for some $x,y\in X$, the definition of $\succ_{c,\psi}$ implies that there is $A\in\X$ such that $x=c(A)$, and $y\in\psi(A)\setminus x$.
Note that $\succ_{c,\psi}$ is irreflexive by definition, and thus $x,y$ must be distinct.
Since $c(A)=\max(\psi(A),\succ_{c})$, and it is unique, we conclude that $\neg(y\succ_{c,\psi} x)$.

($\Longleftarrow$). 
The definition of $\succ_{c,\psi}$ yields $c(A)\succ_{c}x$ for any $A\in \X$ and $x\in\psi(A)\setminus c(A)$.
Since $\succ_{c}$ is asymmetric, we conclude that $c(A)=	\max(\psi(A),\succ_{c,\psi})$. 	
\end{proof}

\begin{lemma}\label{LEM:revealed_preference_contained}
	Let $c\colon \X\to X$ and $\psi\colon \X \to \X$ be respectively a choice function and a choice correspondence on $X$.
	If there is a linear order $\rhd$ on $X$ such that, for any $A\in \X$, we have  $c(A)=\max(\psi(A),\rhd)$, then $\succ_{c,\psi}\,\subseteq \rhd$. 
\end{lemma}
\begin{proof}
	Apply the definition of $\succ_{c,\psi}$.
\end{proof}
\begin{lemma}\label{LEM:linear_order_containing}
	Let $c\colon \X\to X$ and $\psi\colon \X \to \X$ be respectively a choice function and a choice correspondence on $X$.
If there is a linear order $\rhd$ on $X$ such that $\succ_{c,\psi}\,\subseteq \rhd$, then $\max(\psi(A),\succ_{c,\psi})=c(A)=\max(\psi(A),\rhd)$ for any $A\in\X$. 
\end{lemma}
\begin{proof}
	Since $\succ_{c,\psi}\,\subseteq \rhd$, $\succ_{c,\psi}$ is asymmetric, and Lemma~\ref{LEM:asymmetry_revealed_preference} implies that $c(A)=\max(\psi(A),\succ_c)$ for any $A\in \X$. 
	Since $\rhd \supseteq \; \succ_{c,\psi}$, it is a linear order, and $c(A)\succ_{c,\psi} x$ for any $x\in \psi(A)\setminus x$ for any $A\in \X$, we conclude that $c(A)=\max(\psi(A),\rhd)$ for any $A\in\X$.
	\end{proof}
%
%

In \cite{CherepanovFeddersenSandroni2013}, the binary relation $\succ_{c,\psi^{\,\min}}$ is called $Rev$.
Indeed, the authors state that $x\,Rev\,y$ if there are two menus $A,B\in \X$ such that $(A,B)$ is a switch, $y=c(B)$, and $x=c(A)$.
The definition of $\psi^{\,min}$ implies that $x\succ_{c,\psi^{\,min}} y$ if and only if $x\,Rev\,y$.  
Moreover, we have 

\begin{proposition}[\citealt{CherepanovFeddersenSandroni2013}]\label{PROP:rationalization_theorem_characterization}
	A choice function  $c\colon\X\to X$  is consistent with basic rationalization theory if and only if Weak WARP holds.
	A choice function $c\colon\X\to X$ is consistent with ordered rationalization theory if and only if $\succ_{c,\psi^{\,min}}$ is acyclic.  
\end{proposition}
To conclude the proof of Theorem \ref{THM:equivalence_minimal_rationalization_theory}, we need only some additional results.
\begin{lemma}\label{LEM:Weak_WARP_binary_choice}
	Let $c\colon\X\to X$ be a choice function on $X$.
	If $c$ satisfies Weak WARP, and $x\succ_{c,\psi^{\,min}}y$, then $y\neq c(xy)$.
	 \end{lemma}
\begin{proof}
	Assume that $c\colon\X\to X$ satisfies Weak WARP, and $x\succ_{c,\psi^{\,min}}y$ for some $x,y\in\X$. 
	The definition of $\succ_{c,\psi^{\,min}}$ implies that there 
	are $A,B\in\X$ and $x,y\in A$ such that $B\supseteq A$, $x=c(A)$, and $y=c(B)$.
	Weak WARP implies that $y\neq c(xy).$
	\end{proof}
\begin{corollary}\label{COR:Weak_WARP_asymmetry}
	Let $c\colon\X\to X$ be a choice function on $X$.
	If $c$ satisfies Weak WARP, then $\succ_{c,\psi^{\,min}}$ is asymmetric.
\end{corollary}
\begin{lemma}\label{LEM:asymmetry_maximization_Weak_WARP}
Let $c\colon\X\to X$ be a choice function on $X$.
If $\succ_{c,\psi^{\,min}}$ is asymmetric, and for any $A\in\X$ we have $c(A)=\max(\psi^{\,min}(A),\succ_{c,\psi^{\,min}})$, then Weak WARP holds.
\end{lemma}
\begin{proof}
	Let $c\colon\X\to X$ be a choice function on $X$.
Assume that $\succ_{c,\psi^{\,min}}$ is asymmetric, and for any $A\in\X$ we have $c(A)=\max(\psi^{\,min}(A),\succ_{c,\psi^{\,min}})$.
Assume toward a contradiction that Weak WARP fails.
Thus, there are $A,B\in\X$, and $x,y\in B$ such that $B\supseteq A$, $x=c(B)$, $x=c(xy)$, and $y=c(A)$.
Definition~\ref{DEF:minimal_psychological_constraint} and that of $\succ_{c,\psi^{\,min}}$ respectively yield $y\in\psi^{\,min}(A)$ and $y\succ_{c,\psi^{\,min}} x$.
Since $\succ_{c,\psi^{\,min}}$ is asymmetric, we have that $\neg(x\succ_{c,\psi^{\,min}}y)$.
Since $\psi^{\,min}$ satisfies Axiom $\alpha$, we have that $y\in\psi^{\,min}(xy)$.
Since $c(A)=\max(\psi^{\,min}(A),\succ_{c,\psi^{\,min}})$, we obtain that $c(xy)=y$, which is false.
\end{proof}

Finally, we have:

\begin{theorem}\label{THM:Weak_WARP_mbr_Rev_acyclic_mor}
	A choice function $c\colon\X\to X$ satisfies Weak WARP if and only if $\succ_{c,\psi^{\,min}}$ is asymmetric, and $c(A)=\max(\psi^{\,min}(A),\succ_{c,\psi^{\,min}})$ for any $A\in\X$.
	Moreover, $\succ_{c,\psi^{\,min}}$ is acyclic if and only if there is a linear order $\rhd$ on $X$ including $\succ_{c,\psi^{\,min}}$ such that $c(A)=\max(\psi^{\,min}(A),\rhd)$ for any $A\in\X$.
\end{theorem}
\begin{proof}
	Let $c\colon\X\to X$ be a choice function on $X$.  
	
	($\Longrightarrow$). Assume that $c$ satisfies Weak WARP. Corollary~\ref{COR:Weak_WARP_asymmetry} and Lemma~\ref{LEM:asymmetry_revealed_preference} yield that $\succ_{c,\psi^{\,min}}$ is asymmetric, and $c(A)=\max(\psi^{\,min}(A),\succ_{c,\psi^{\,min}})$ for any $A\in\X$.
Assume now that Rev is acyclic.
By \cite{Szpilrajn1930}'s theorem there is a linear order $\rhd$ on $X$ that includes $\succ_{c,\psi^{\,min}}$.
Lemma~\ref{LEM:linear_order_containing} yields $c(A)=\max(\psi^{\,min}(A),\succ_{c,\psi^{\,min}})$ for any $A\in\X$.
	
	($\Longleftarrow$). Assume that $\succ_{c,\psi^{\,min}}$ is asymmetric, and $c(A)=\max(\psi^{\,min}(A),\succ_{c,\psi^{\,min}})$ for any $A\in\X$.
	By Lemma~\ref{LEM:asymmetry_maximization_Weak_WARP} Weak WARP holds.
	Assume now that there is a linear order $\rhd$ on $X$ including $\succ_{c,\psi^{\,min}}$ such that $c(A)=\max(\psi^{\,min}(A),\rhd)$ for any $A\in\X$.
	Since $\rhd\supseteq\, \succ_{c,\psi^{\,min}} $, we conclude that $\succ_{c,\psi^{\,min}} $ is acyclic.
	\end{proof}

Theorem~\ref{THM:equivalence_minimal_rationalization_theory} is an immediate corollary of  Proposition~\ref{PROP:rationalization_theorem_characterization} and Theorem~\ref{THM:Weak_WARP_mbr_Rev_acyclic_mor}.
\qed

\medskip

\noindent \underline{\it Proof of Lemma~\ref{LEM:minimal_psychological_constraint_including_minimal_attention_filter}.}
Let $c\colon\X\to X$ be a choice function on $X$.
Let $\psi\colon\X\to \X$ be a choice correspondence on $X$ satisfying Axiom $\alpha$ such that $\psi(A)\supseteq \Gamma^{\,min}(A)$ for any $A\in \X$.

We first show that Axiom $\alpha$ holds for $\psi^{\,min}_{*}$.
For some menu $A\in\X$, take $x\in\psi^{\,min}_{*}(A)$.
The definition of $\psi^{\,min}_{*}$ implies that either $x\in S_B$ or $x\in D_B$ for some $B\supseteq A$.
Let  $C\subseteq A$ be a menu containing $x$.
Since $C\subseteq A$,  either $x\in S_B$ or $x\in D_B$ for some $B\supseteq C$.
The definition of $\psi^{\,min}_{*}$ implies that $x\in\psi^{\,min}_{*}(C)$.

To show that $\psi^{\,min}_{*}(A)\subseteq \psi(A)$ for any $A\in\X$, first note that by definition $\psi_{*}^{\,min}(A)=\bigcup_{B\supseteq A}\Gamma^{\,min}(B)\cap A$ holds.
Since $\psi(A)\supseteq  \Gamma^{\min}(A)$ for any $A\in \X$, and Axiom $\alpha$ holds for $\psi$, we have that $\psi^{\,min}(A)\subseteq \bigcup_{B\supseteq A}\psi(B)\cap A \subseteq \psi (A)$.\qed
\medskip

\noindent\underline{\it Proof of Theorem \ref{THM:minimally_attentive_rationalization_theory_characterization}.}
Let $c\colon\X\to X$ be a choice function on $X$.

($\Longrightarrow$).
Assume that $c$ is consistent with minimally attentive basic rationalization theory, and let $(\psi^{\,min}_{*},\succ)$ be some explanation by limited consideration of $c$.
Toward a contradiction, assume that that $x R y$ and $y R x$, for some $x,y\in X$.
Since $x R y$, Definition~\ref{DEF:revealed_preference_minimally_attentive_rationalization theory} implies that there are $x,y\in X$ and $A,B\in\X$ such that $B\supseteq A$, $x,y\in A$, $x=c(A)$,and $y\in S_A^{\uparrow}\cup D_A$.
Since $\psi^{\,min}_{*}$ satisfies Axiom $\alpha$ and $y\in\psi^{\,min}_{*}(B)\cap A$, we have that $y\in\psi^{\,min}_{*}(A)$.
Since $x=c(A)=\max(\psi^{\,min}_{*},\succ)$  and  $y\in\psi^{\,min}_{*}(A)$, we conclude that $x\succ y$.
Since $y R x$, Definition~\ref{DEF:revealed_preference_minimally_attentive_rationalization theory} implies that there are $C,D\in\X$ such that $C\supseteq D$, $x,y\in D$, $y=c(D)$, and $x\in S_C^{\uparrow}\cup D_C$.
Axiom $\alpha$ implies that $x\in \psi^{\,min}_{*}(D)$, which is impossible.

Assume now that $c$ is consistent with minimally attentive ordered rationalization theory, and let $(\psi^{\,min}_{*},\rhd)$ be some explanation by limited consideration of $c$. 
Toward a contradiction, assume that there are $x_1,x_2\cdots,x_n \in X$ such that $x_1 R\,x_2 \,R\cdots R\,x_n R\, x_1 $, for some $x,y\in X$.
 Since $x_1 \,R \,x_2$, Definition~\ref{DEF:revealed_preference_minimally_attentive_rationalization theory} implies that there are $A,B\in\X$ such that $B\supseteq A$, $x,y\in A$, $x=c(A)$, and $y\in S_A^{\uparrow}\cup D_A$.
Since $\psi^{\,min}_{*}$ satisfies Axiom $\alpha$ and $y\in\psi^{\,min}_{*}(B)\cap A$, we have that $x_2\in\psi^{\,min}_{*}(A)$.
Since $x_1=c(A)=\max(\psi^{\,min}_{*},\rhd)$  and  $x_2\in\psi^{\,min}_{*}(A)$, we have that $x_1\rhd x_2$.
Using the same argument we conclude that $x_i\rhd x_{i+1}$ for any $i\leq n$, and $x_n\rhd x_1$.
Thus, we have $x_1 \rhd x_2\rhd \cdots x_n \rhd x_1$, which is impossible, since $\rhd$ is a linear order.

 ($\Longleftarrow$).
Assume that $P$ is asymmetric.
Observe that $c(A)P y$ for any $A\in \X$, and $y\in \psi^{\,min}_{*}(A)\setminus c(A)$.
Since $P$ is asymmetric, we conclude that $c(A)=\max(\psi^{\,min}_{*}(A),P)$ for any $A\in\X$

Assume that $P$ is asymmetric and acyclic.
Observe that $c(A)P y$ for any $A\in \X$, and $y\in \psi^{\,min}_{*}(A)\setminus c(A)$.
Since $P$ is asymmetric and acyclic, by \cite{Szpilrajn1930}'s theorem there is a linear order $\rhd$ such that $\rhd \supseteq P$.
 We conclude that $c(A)=\max(\psi^{\,min}_{*}(A),\rhd)$ for any $A\in\X$.\qed
 

\begin{thebibliography}{31}

\providecommand{\natexlab}[1]{#1}
\providecommand{\url}[1]{\texttt{#1}}
\expandafter\ifx\csname urlstyle\endcsname\relax 
  \providecommand{\doi}[1]{doi: #1}\else
  \providecommand{\doi}{doi: \begingroup \urlstyle{rm}\Url}\fi


\bibitem[Abaluck and Adams-Prassl(2021)]{AbaluckAdams-Prassl}{\textsc{Abaluck, J., and Adams-Prassl, A.}, 2021.
What do consumers consider before they choose? Identification from asymmetric demand responses.
\textit{The Quarterly Journal of Economics} 136:\,1611--1663.}


\bibitem[Aguiar et al.(2023)]{Aguiaretal2023}{\textsc{Aguiar, V.\,H., Boccardi, M.\,J., Kashaev, N., and Kim, J.}, 2023.
Random utility and limited consideration.
\textit{Quantitative Economics} 14:\,71--116.}



\bibitem[Allenby and Ginter(1995)]{AllenbyGinter1995}{\textsc{Allenby, G., and Ginter, J.}, 1995.
The effects of in-store displays and feature advertising on consideration sets.
\textit{Journal of Research in Marketing} 12:\,67--80.}




%

%
\bibitem[Arrow(1963)]{Arrow1963}
{\textsc{Arrow, K.}, 1963.
\textit{Social choice and individual values.} Second Edition. New York: Wiley.}




\bibitem[Bordalo, Gennaioli, and Shleifer(2012)]{BordaloGennaioliShleifer2012}{\textsc{Bordalo, P., Gennaioli, N., and Shleifer, A.}, 2012.
Salience theory of choice under risk. 
\textit{Quarterly Journal of Economics} 127:\,1243--1285.}

\bibitem[Bordalo, Gennaioli, and Shleifer(2013)]{BordaloGennaioliShleifer2013}{\textsc{Bordalo, P., Gennaioli, N., and Shleifer, A.}, 2013.
Salience and consumer choice. 
\textit{Journal of Political Economy} 121:\,803--843.}


\bibitem[Brady and Rehbeck(2016)]{BradyRehbeck2016}{\textsc{Brady, R., and Rehbeck}, 2016.
Menu-dependent stochastic feasibility.
\textit{Econometrica} 84:\,1203--1223.}

\bibitem[Cantone et al.(2016)]{CantoneGiarlottaGrecoWatson2016}{\textsc{Cantone, D., Giarlotta, A., Greco, S., and Watson, S.}, 2016.
$(m,n)$-rationalizable choices.
\textit{Journal of Mathematical Psychology}
73:\,12--27.}

\bibitem[Caplin and Dean(2015)]{CaplinDean2015}{\textsc{Caplin, A., and Dean, M.}, 2015.
Revealed preference, rational inattention, and costly information acquisition.
\textit{American Economic Review}
105:\,2183--2203.}



\bibitem[Caplin, Dean, and Leahy(2019)]{CaplinDeanLeahy2019}{\textsc{Caplin, A., Dean, M., and Leahy, J.}, 2019.
Rational inattention, optimal consideration sets, and stochastic choice.
\textit{The Review of Economic Studies} 86:\,1061--1094.}


\bibitem[Cattaneo et al.(2020)]{Cattaneoetal2020}{\textsc{Cattaneo, M.\, D., Ma, X., Masatlioglu, Y., and Suleymanov, E.}, 2020.
A random attention model.
\textit{Journal of Political Economy} 128:\,2796--2836.
}

%


\bibitem[Cherepanov, Feddersen, and Sandroni(2013)]{CherepanovFeddersenSandroni2013}
{\textsc{Cherepanov, V., Feddersen, T., and Sandroni, A.}, 2013.
Rationalization. 
\textit{Theoretical Economics} 8:\,775--800.}

\bibitem[Chernoff(1954)]{Chernoff1954}
{\textsc{Chernoff, H.}, 1954. 
Rational selection of decision functions. 
\textit{Econometrica} 22:\,422--443.}


\bibitem[Chiang, Ching, and Narasimhan(1999)]{ChiangChibNarasimhan1999}{\textsc{Chiang, J., Chib, S., and Narasimhan, C.}, 1999.
Markov chain Monte Carlo and models of consideration set and parameter heterogeneity.
\textit{Journal of Econometrics} 89:\,223--248.}


\bibitem[Crompton and Ankomah(1993)]{CromptonAnkomah1993}{\textsc{Crompton, J,\,L., and Ankomah, P.\,K.}, 1993.
Choice set propositions in destinations decisions.
\textit{Annual Tourism Research} 20:\,461--476.}


%
 



\bibitem[de Clippel and Rozen(2023)]{DeClippelRozen2023}{\textsc{de Clippel, G., and Rozen, K.}, 2023.
Bounded rationality in choice theory: a survey.
\textit{Mimeo, commissioned by the JEL}. }

\bibitem[Demuynck and Seel(2018)]{DemuynckSeel2018}{\textsc{Demuynck, T., and Seel, C.}, 2018.
Revealed preference with limited consideration.
\textit{American Economic Journal: Microeconomics} 10:\,102--131.}



\bibitem[Eliaz and Spiegler(2011)]{EliazSpiegler2011}{\textsc{Eliaz, K., and Spiegler, R.}, Consideration sets and competitive marketing. 
\textsl{The Review of Economic Studies} 78:\,235--262.}


\bibitem[Fader and McAlister(1990)]{FaderMcAlister1990}{\textsc{Fader, P., and McAlister, L.}, 1990.
An elimination by aspects model of consumer response to promotion calibrated on UPC scanner data.
\textit{Journal of Marketing Research} 27:\,322--332.}

\bibitem[Geng and Ozbay(2021)]{GengOzbay2021}{\textsc{Geng, S., and Ozbay, E.}, 2021. Shortlisting procedure with a limited capacity.
\textit{Journal of Mathematical Economics} 94.}

\bibitem[Geng(2022)]{Geng2022}{\textsc{Geng, S.}, 2022.
Limited consideration model with a trigger or a capacity.
\textit{Journal of Mathematical Economics} 101.}

\bibitem[Giarlotta, Petralia, and Watson(2022)]{GiarlottaPetraliaWatson2022a}{\textsc{Giarlotta, A., Petralia, A., and Watson, S.}, 2022.
Bounded rationality is rare.
\textit{Journal of Economic Theory} 204, 105509.} 

\bibitem[Giarlotta, Petralia, and Watson(2023)]{GiarlottaPetraliaWatson2022b}
{\textsc{Giarlotta, A., Petralia, A., and Watson, S.}, 2023.
Context-Sensitive Rationality: Choice by Salience.
\textit{Journal of Mathematical Economics} 109, 102913.}


\bibitem[Gilbride and Allenby(2004)]{GilbrideAllenby2004}{\textsc{Gilbride, T., and Allenby, G.}, 2004.
A choice model with conjunctive, disjunctive, and compensatory screening rules.
\textit{Marketing Science} 23:\,391--406.}







%

\bibitem[Hauser and Wernerfelt(1990)]{HauserWernerfelt1990}{\textsc{Hauser, J., R., and Wernerfelt, B.}, 1990.
An Evaluation Cost Model of Consideration Sets.
\textit{Journal of Consumer Research} 16:\,393--408.}












\bibitem[Kibris, Masatlioglu, and Suleymanov(2021)]{KibrisMasatliogluSuleymanov2021}
{\textsc{Kibris, O., Masatlioglu, Y., and Suleymanov, E.}, 2021.
A Theory of Reference Point Formation.
\textit{Economic Theory}.}





\bibitem[Lanzani(2022)]{Lanzani2022}{\textsc{Lanzani, G.}, 2022.
Correlation made simple: application to salience and regret theory.
\textit{The Quaterly Journal of Economics} 137:\,959--987.}


\bibitem[Lim(2021)]{Lim2021}{\textsc{Lim, X.\,Z.}, 2021.
Ordered Reference Dependent Choice. arXiv:2105.12915 [econ.TH].}

\bibitem[Lleras, Masatlioglu, Nakajima, and Ozbay(2017)]{LlerasMasatliogluNakajimaOzbay2017}
{\textsc{Lleras, J.,\, S, Masatlioglu, Y., Nakajima, D., and Ozbay, E.\,Y.}, 2017.
When more is less: limited consideration.
\textit{Journal of Economic Theory} 170:\,70--85.} 



%




%
\bibitem [Manzini and Mariotti(2012)]{ManziniMariotti2012b}
{\textsc{Manzini, P., and Mariotti, M.}, 2012.
Categorize then choose: boundedly rational choice and welfare.  
\textit{Journal of the European Economic Association} 10:\,1141--1165.}
%

\bibitem[Manzini and Mariotti(2014)]{ManziniMariotti2014}{\textsc{Manzini, P., and Mariotti, M.}, (2014).
Stochastic choice and consideration sets.
\textit{Econometrica} 82:\,1153-1176.}

%
\bibitem[Masatlioglu, Nakajima, and Ozbay(2012)]{MasatliogluNakajimaOzbay2012}
{\textsc{Masatlioglu, Y., Nakajima, D., and Ozbay, E.\,Y.}, 2012.
Revealed attention. 
\textit{American Economic Review} 102:\,2183--2205.}


\bibitem[Mat\u{e}jka, and Mckay(2015)]{MatejkaMckay2015}{\textsc{Mat\u{e}jka, F., and McKay, A.}, 2015.
Rational inattention to discrete choices: a new foundation for the multinomial logit model.
\textit{American Economic Review} 105:\, 272--298.}



\bibitem[Meissner, Strauss, and Talluri(2013)]{MeissnerStraussTalluri2013}{\textsc{Meissner, J., Strauss, A., and Talluri, K.}, 2013.
Enhanced concave program relaxation for choice network revenue management.
\textit{Production Operational Management} 22:\,71--87.}


%

\bibitem[Parr and Friston(2017)]{ParrandFriston2017}{\textsc{Parr, T., and Friston, K.\,J.} 2017.
Working memory, attention, and salience in active inference.
\textit{Scientific Reports} 7:\,14678.}

\bibitem[Parr and Friston(2019)]{ParrandFriston2019}{\textsc{Parr, T., and Friston, K.\,J.} 2019.
Attention or Salience?
\textit{Current Opinion in Psychology} 29:\,1--5.}


\bibitem[Paulssen and Bagozzi(2005)]{PaulssenBagozzi2005}{\textsc{Paulssen, M, and Bagozzi, R.\,P.}, 2005.
A self-regulatory model of consideration set formation.
\textit{Psychology and \& Marketing} 22:\,785--812.}


%







\bibitem[Ravid and Stevenson(2021)]{RavidStevenson2021}{\textsc{Ravid, D., and Stevenson, K.}, 2021.
Bad temptation.
\textit{Journal of Mathematical Economics} 95:\,102480.}


\bibitem[Roberts and Lattin(1991)]{RobertsLattin1991}{\textsc{Roberts, J., and Lattin, J.}, 1991.
Development and testing of a model of consideration set composition.
\textit{Journal of Marketing Research} 28:\,429--440.}

%

\bibitem[Samuelson(1938)]{Samuelson1938}
{\textsc{Samuelson, A.\,P.}, 1938.
A note on the pure theory of consumer's behaviour.
\textit{Economica} 17:\,61--71.} 
%

\bibitem[Sen(1971)]{Sen1971}{\textsc{Sen, A.\,K.}, 1971.
Choice functions and revealed preference. 
\textit{The Review of Economic Studies} 38:\,307--317.}

\bibitem[Sen(1986)]{Sen1986}{\textsc{Sen, A.\,K.}, 1986.
Social choice theory.
In K.\,J. Arrow and \& M.\,D. Intrilligator (Eds.),
\textit{Handbook of Mathematical Economics Vol III} (pp. 1073--1081). Elsevier Science Publisher, North-Holland.}

\bibitem[Simon(1955)]{Simon1955}
{\textsc{Simon, H.\,A.}, 1955.
A behavioral model of rational choice.
\textit{Quarterly Journal of Economics} 69:\,99--118.}

%


\bibitem[Sims(2003)]{Sims2003}{\textsc{Sims, C.}, 2003.
Implications of rational inattention.
\textit{Journal of Monetary Economics} 50:\,665--690.}





\bibitem[Suh(2009)]{Suh2009}{\textsc{Suh, J.\,C.}, 2009.
The role of consideration sets in brand choice: The moderating role of product characteristics.
\textit{Psychology \& Marketing}, 26:\,534--550.}


\bibitem[Szpilrajn(1930)]{Szpilrajn1930}{\textsc{Szpilrajn, E.}, 1930.
Sur l'extension de l'ordre partiel. 
\textit{Fundamenta Mathematicae} 16: 386--389.}

\bibitem[Terui, Ban, and Allenby(2011)]{TeruiBanAllenby2021}{\textsc{Terui, N., Ban, M., and Allenby, G.\,M.}, 2011.
Advertising on brand consideration and choice.
\textit{Marketing Science} 30:\,74--91.}


\bibitem[Van Nierop et al.(2010)]{VanNieropetal}{\textsc{Van Nierop, E., Bronnenberg, B., Paap, R., Wedel, M., and Franses, P.\,H.}, 2010.
Retrieving unobserved consideration sets from household panel data.
\textit{Journal of Marketing Research}, 47:\,63--74.}


\bibitem[Yildiz(2016)]{Yildiz2016}
{\textsc{Yildiz, K.}, 2016.
List-rationalizable choice.
\textit{Theoretical Economics} 11:\,587--599.}




\end{thebibliography}
\end{document}